\documentclass[journal=jpcafh,manuscript=article]{achemso}
\usepackage[utf8]{inputenc}

\usepackage{longtable}
\usepackage{multirow}
\usepackage{amsmath}
\usepackage{subfigure}
\usepackage[usenames,dvipsnames]{xcolor}
\usepackage{soul}
\usepackage{bm}

\usepackage{xr}
\usepackage{amssymb}
\usepackage{siunitx}
\usepackage{threeparttable}
\usepackage{multicol} \usepackage{wrapfig} \usepackage{enumitem}
\usepackage{bm} \usepackage{outlines} 
\usepackage{booktabs}
\usepackage{graphicx}
\usepackage[normalem]{ulem}
\usepackage{hyperref}
\usepackage{lineno}
\usepackage[version=4]{mhchem}
\usepackage[section]{placeins}

\makeatletter
\newcommand*{\addFileDependency}[1]{
  \typeout{(#1)}
  \@addtofilelist{#1}
  \IfFileExists{#1}{}{\typeout{No file #1.}}
}
\makeatother

\newcommand*{\myexternaldocument}[1]{%
    \externaldocument{#1}%
    \addFileDependency{#1.tex}%
    \addFileDependency{#1.aux}%
}
\myexternaldocument{si}

\author{Cangtao Yin, Stefan Willitsch} \affiliation{Department of Chemistry,
  University of Basel, Klingelbergstrasse 80, CH-4056 Basel,
  Switzerland}

\author{Markus Meuwly}\email{m.meuwly@unibas.ch}
\affiliation{Department of Chemistry, University of Basel,
  Klingelbergstrasse 80, CH-4056 Basel, Switzerland}

\title[]{Full-Dimensional Reactive Potential Energy Surfaces for
  OCS$^+$ $\rightarrow$ CO+S$^+$ Dissociation: Ground and Excited
  States}

\begin{document}
\date{\today}

\begin{abstract}
Full-dimensional reactive potential energy surfaces (PESs) for the
OCS$^+$ cation are constructed to describe S$^+$ loss in the
electronic ground state and seven low-lying electronically excited
states. High-level \textit{ab initio} reference energies were computed
at the MRCI+Q/aug-cc-pVTZ level and were used to generate PESs
employing reproducing kernel Hilbert space representations (RKHS). The
PESs accurately reproduce the measured dissociation limits to
CO(X$^1\Sigma^+$)+S$^+$ in different electronic states. The topology
of the PESs reveals multiple linear and T-shaped minima, pronounced
angular anisotropy, and state-crossing manifolds. Exploratory
quasi-classical trajectory simulations on selected PESs confirm
numerical stability and energy conservation, illustrating the
suitability of the surfaces for dynamical applications. The present
work represents the most comprehensive characterization to date of the
lowest PESs of OCS$^+$ and provides a reliable foundation for future
studies of the photodissociation of OCS$^+$ and the chem-ionization
dynamics of OCS.
\end{abstract}

\section{Introduction}
Carbonyl sulfide OCS is a prototypical linear triatomic molecule and a
major sulfur-containing constituent of planetary
atmospheres\cite{ma21a} and the interstellar medium.\cite{matthews87a}
The ionization of OCS and the structure and dynamics of its cation
OCS$^+$ have been the subject of a variety of studies in the realms of
photo-ionization and photoelectron
spectroscopy\cite{eland:1973,wang88a,holland16a,wiese19a,ramadhan16a},
photodissociation dynamics\cite{chang05a,wang24a,wang24b} as well as
chemi- and electron-impact
ionization\cite{kishimoto03a,horio06a,ploenes2021novel,lomas25a}. These
experimental studies revealed complex electronic and nuclear dynamics
in the cation which is governed by the close proximity and interaction
of several low-lying electronic states. Therefore, a detailed
understanding of the involved processes necessitates a unified
description of the reactive potential energy surfaces (PESs) of the
lowest states of OCS$^+$.\\

\noindent
This system was previously investigated following 1-dimensional scans
along the collinear dissociation.\cite{hirst:2006,chen:2006} These
studies were carried out at the MRCI level with the cc-pVTZ basis set
using a (13e,7o) active space\cite{hirst:2006}, and at the CASPT2
level of theory with an ANO-type basis set using a (11e,13o) active
space\cite{chen:2006}. Their calculations indicate that the S$^+$-loss
dissociation products of the OCS$^+$ ion are the ground-state CO
molecule plus the S$^+$ ion in different electronic
states. Subsequently, CASPT2 potential energy curves for O-loss
dissociation from several low-lying states were reported in
Ref. \cite{dong2011loss}. More recently, the dissociative electron
ionization dynamics of OCS were investigated using EOM-CCSD
calculations for both the OCS$^+$ cation and the OCS$^{2+}$ dication
\cite{lomas25a}. In that work, one-dimensional potential energy curves
were computed for both S-loss and O-loss dissociation pathways. The
goal in this work is to obtain a full-dimensional PES suitable for
reactive molecular dynamics simulations.\\

\noindent
The present work is structured as follows. First, the electronic
structure method, means to represent the reactive PESs, and the quasi
classical trajectory simulations are described. This is followed by
results on the quality of the PESs, their features and
interrelationships. Quasi-classical trajectory (QCT) simulations
provide a means to validate the PESs, which is described
next. Finally, conclusions are drawn.\\

\section{Methods}

\subsection{Electronic Structure Calculations}
The reactive PESs to describe S$^+$ dissociation from OCS$^+$ in the
electronic ground and seven excited states were
constructed at the MRCI+Q\cite{Werner:1988} level of theory together
with the aug-cc-pVTZ (aVTZ)\cite{Dunning:1989} basis set. All
electronic structure calculations were performed in Jacobi coordinates
$(R, r, \theta)$ and in $C_{\rm S}$ symmetry using the Molpro software
package.\cite{MOLPRO} Here, $R$
represents the distance between S$^+$ and the center of mass of CO,
$r$ is the CO bond length, and $\theta$ is the angle between the
vectors $R$ and $r$.\\

\noindent
The reference data were computed on a three-dimensional grid defined
by $R \in$ [2.35, 2.45, 2.54, 2.64, 2.73, 2.83, 2.92, 3.02, 3.11,
  3.20, 3.30, 3.39, 3.49, 3.58, 3.68, 3.77, 3.87, 3.96, 4.06, 4.15,
  4.34, 4.53, 4.72, 4.91, 5.09, 5.28, 5.47, 5.66, 5.85, 6.04, 6.23,
  6.42, 7.17, 8.12, 9.06, 10.01, 10.95] $a_0$, $r \in$ [1.97, 2.01,
  2.06, 2.15, 2.24, 2.32, 2.38] $a_0$, and $\theta \in$ [11.98, 27.49,
  43.10, 58.73, 74.36, 90.00, 105.64, 121.27, 136.90, 152.51,
  168.02]$^\circ$. The $r-$grid includes the classical turning points
of the CO-vibration which were determined by solving the 1-dimensional
Schr\"odinger equation using the LEVEL program.\cite{level} Along the
$\theta-$coordinate, a Legendre-grid was chosen,\cite{MM.heh2:1999}
which is useful when solving the 3-dimensional nuclear Schr\"odinger
equation in future work to determine, e.g., accurate bound state
energies.  Overall, the grid consists of 2849 geometries.\\

\noindent
For each geometry, CASSCF calculations were performed with equal
weighting of all states, using an active space of 11 electrons in 13
orbitals, followed by MRCI calculations with an active space of 11
electrons in 10 orbitals. The MRCI+Q approach was chosen over standard
MRCI to improve accuracy through the Davidson relaxed reference
correction.\cite{Langhoff:1974} As is common in such high-level
electronic structure calculations, some CASSCF or MRCI computations
either failed to converge or converged to an incorrect electronic
state. These problematic cases, along with configurations exhibiting
total energies $> 10$ eV above the global minimum were excluded. The
resulting cleaned dataset was then used to construct a
three-dimensional reproducing kernel Hilbert space (RKHS)
representation\cite{MM.rkhs:2017} for each electronic state.\\

\noindent
For validating the RKHS-PESs, an independent “off-grid” dataset was
generated. These geometries were defined by $R$ = [3.0, 4.0, 5.0, 6.0,
  7.0, 8.0, 9.0, 10.0, 11.0, 12.0] $a_0$, $r$ = [2.0, 2.1, 2.2, 2.3]
$a_0$, and $\theta = [30.0, 60.0, 90.0, 120.0, 150.0]^\circ$. None of
these off-grid geometries were included in the RKHS-representation of
the PESs. All corresponding reference energies were computed again at
the MRCI+Q/aVTZ level, and as with the on-grid data, unconverged and
high-energy points (more than 10 eV above the global minimum) were
removed.\\

\subsection{Representation of the PESs}
All PESs in the present work were represented as a RKHS. To ensure a
correct description of the S$^+$+CO dissociation limit, the total
energy of OCS$^+$ was written as $V(R,r,\theta) = E(R,r,\theta) +
V(r)$, where $V(r)$ denotes the diatomic potential of CO. The
\textit{ab initio} potential energy curve of CO was computed
independently and represented using a one-dimensional RKHS with a
reciprocal-power decay kernel
$k^{[2,4]}(r,r')$.\cite{rabitz:1996,MM.rkhs:2017} The total
interaction energy $E(R,r,\theta)$ was shifted such that
$E(R,r,\theta) \rightarrow 0$ as $R \rightarrow \infty$. In the
construction of the three-dimensional PESs, reciprocal power kernels
$k^{[n,m]}(x_<,x_>)$ with parameters $n=2$ and $m=4$ were employed for
the radial coordinates $R$ and $r$, while a Taylor spline kernel
($n=2$) was used for the coordinate $z$, which is related to the
angular coordinate $\theta$ by $z=(1-\cos{\theta})/2$.\\

\subsection{QCT Simulations}
To validate the RKHS representation of the PESs, QCT simulations were
performed.\cite{koner:2016,MM.cno:2018} The QCT method used in this
work has been thoroughly described in the
literature.\cite{karplus:1965,truhlar:1979,henriksen2008reaction_dynamics,koner:2016,MM.cno:2018,MM.no2:2020,MM.co2:2021}
A total of 10000 independent trajectories were propagated on each of
the $1 ^2{\rm A}''$ and $1 ^4{\rm A}''$ PESs. Hamilton’s coupled
differential equations of motion were integrated in reactant Jacobi
coordinates using the fourth-order Runge–Kutta method with a time step
of $\Delta t = 0.05$~fs. The trajectories were terminated when the
separation between CO and S$^+$ exceeded $40\,a_0$. Given the rather
different overall topologies of the two PESs considered in the QCT
simulations - $1 ^2{\rm A}''$ features a deep well whereas $1 ^4{\rm
  A}''$ is primarily dissociative - different types of initial
conditions were used.  For the dynamics on the $1 ^2{\rm A}''$ PES,
the initial separation between CO and S$^+$ was set to $30\,a_0$, and
the maximum impact parameter was $b_{\rm max} = 25\,a_0$, with $b$
sampled randomly over the interval $0 \le b \le 25\,a_0$. The initial
collision energy was 0.05~eV. When probing the $1 ^4{\rm A}''$ PES,
the initial conditions were as follows: $R \in [4.0, 4.8]\,a_0$, $r
\in [2.06, 2.24]\,a_0$, and $\theta \in [0, 180^\circ]$. The initial
temperature was set to 5~K (corresponding to 0.00043 eV). In both
cases, the initial vibrational and rotational states are set to the
ground state, $(v=0,j=0)$.\\

\noindent
From the trajectories, the final rotational angular momentum quantum
number $j'$ of the diatomic fragment was determined from the classical
rotational angular momentum evaluated at large intermolecular
separation and subsequently assigned to the nearest integer value of
$j'$. To assess the numerical stability and accuracy of the trajectory
calculations, energy conservation was monitored throughout the
simulations. In particular, the distribution of deviations of the
total energy from its mean value, $P(E- <E>)$, was evaluated,
providing a quantitative measure of the energy conservation achieved
in the calculations. In addition, the opacity function $P(b)$,
describing the probability of reactive complex formation as a function
of the impact parameter $b$, was calculated and discussed.\\

\section{Results and Discussion}

\subsection{1d OC--S$^+$ Scan}
\begin{figure}[H]
    \centering
    \includegraphics[width=1.0\linewidth]{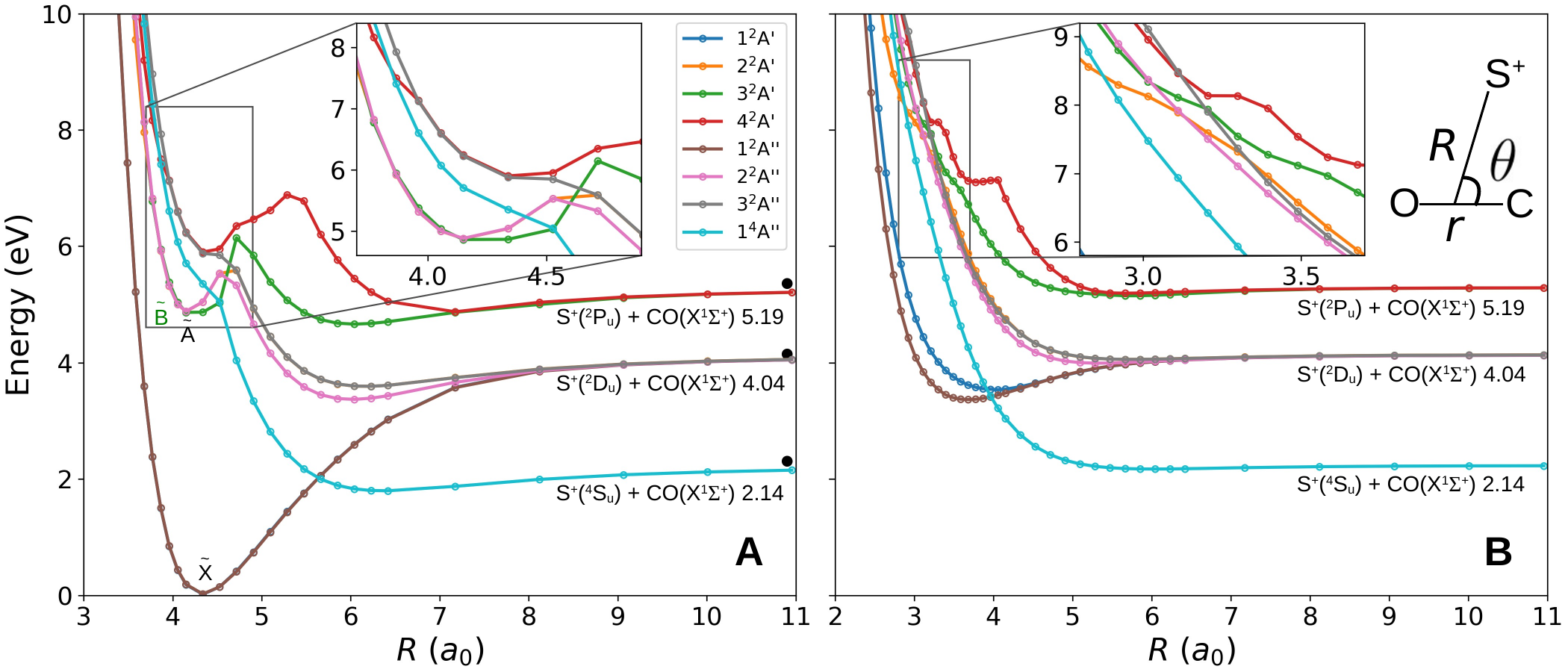}
    \caption{Potential energy curves of the OCS$^+$ system calculated
      at the MRCI+Q/aVTZ level with an active space of 11 electrons in
      10 orbitals. The lines connecting the symbols are to guide the
      eye. The CO-bond length is fixed at $r = 2.15~a_0$,
      corresponding to the equilibrium distance in the OCS$^+$
      molecule. Panel A: $\theta = 1^\circ$, and panel B: $\theta =
      90^\circ$. The experimental energies of the three lowest
      dissociation asymptotes are indicated as black circles for
      comparison \cite{eland:1973}. Insets provide a magnified view of
      the state-crossing regions. The asymptotes and their associated
      dissociation products are presented, along with the electronic
      states identified by their spectroscopic labels. The
      $\tilde{\mathrm{X}}$ state corresponds to the $1^2{\rm A}'$ and
      $1^2{\rm A}''$ states, the $\tilde{\mathrm{A}}$ state
      corresponds to the $2^2{\rm A}'$ and $2^2{\rm A}''$ states, and
      the $\tilde{\mathrm{B}}$ state corresponds to the $3^2{\rm
        A}'$state. The near-degeneracy between the minima of the
      $\tilde{\mathrm{A}}$ and $\tilde{\mathrm{B}}$ states in panel A
      arises because the calculation was performed at a fixed $r =
      2.15~a_0$. When $r$ is allowed to relax, $\tilde{\mathrm{A}}$
      decreases in energy and shows better agreement with experiment
      (see text).}
    \label{fig:pes-1d}
\end{figure}

Figure \ref{fig:pes-1d} reports 1-dimensional scans along the
$R-$coordinate for the nearly linear (panel A) and T-shaped (panel B)
configurations for the ground ($1^2{\rm A}''$) and seven lowest
electronically excited states. An angle $\theta = 1^\circ$ was adopted
to break the linear symmetry in Figure \ref{fig:pes-1d}A, allowing to
use the state notation in the planar ($C_{\rm S}$ symmetry) geometry,
while remaining close enough to linearity to enable comparison with
experimentally measured excitation energies.\cite{eland:1973}. In
these scans, the CO diatomic bond length was fixed at $r = 2.15~a_0$
corresponding to the equilibrium distance in the ground state of
OCS$^+$. \\

\noindent
At long range of $R$, the $1 ^2{\rm A}'$ (blue), $2 ^2{\rm A}'$
(orange), $1 ^2{\rm A}''$ (brown), $2 ^2{\rm A}''$ (magenta), and $3
^2{\rm A}''$ (grey) states are degenerate, see Figure
\ref{fig:pes-1d}. When the system is nearly linear ($\theta =
1^\circ$, Figure \ref{fig:pes-1d}A), the 1d curves for the $1 ^2{\rm
  A}'$ and $1 ^2{\rm A}''$ states are almost degenerate for all values
of $R$. Similarly, the $3 ^2{\rm A}'$ (green) and $4 ^2{\rm A}'$ (red)
states are degenerate for $R\rightarrow \infty$, whereas the $1 ^4{\rm
  A}''$ (light blue) is singly degenerate and asymptotically
corresponds to the lowest energy state.\\

\noindent
Compared to the nearly linear case in Figure \ref{fig:pes-1d}A, the
T-shaped configuration in Figure \ref{fig:pes-1d}B shows a
significantly different picture. The one-dimensional PESs for the $1
^2{\rm A}'$ and $1 ^2{\rm A}''$ states are no longer degenerate,
reflecting the reduced symmetry. The shape of the $1 ^4{\rm A}''$
(light blue) curve remains similar to that in the nearly linear
geometry. For the remaining states, no minima are observed at short
$R$, unlike in the nearly linear case where distinct minima are
present.\\

\subsection{The 3d Reproducing Kernel PESs}
\begin{figure}[H]
    \centering \includegraphics[width=1.0\linewidth]{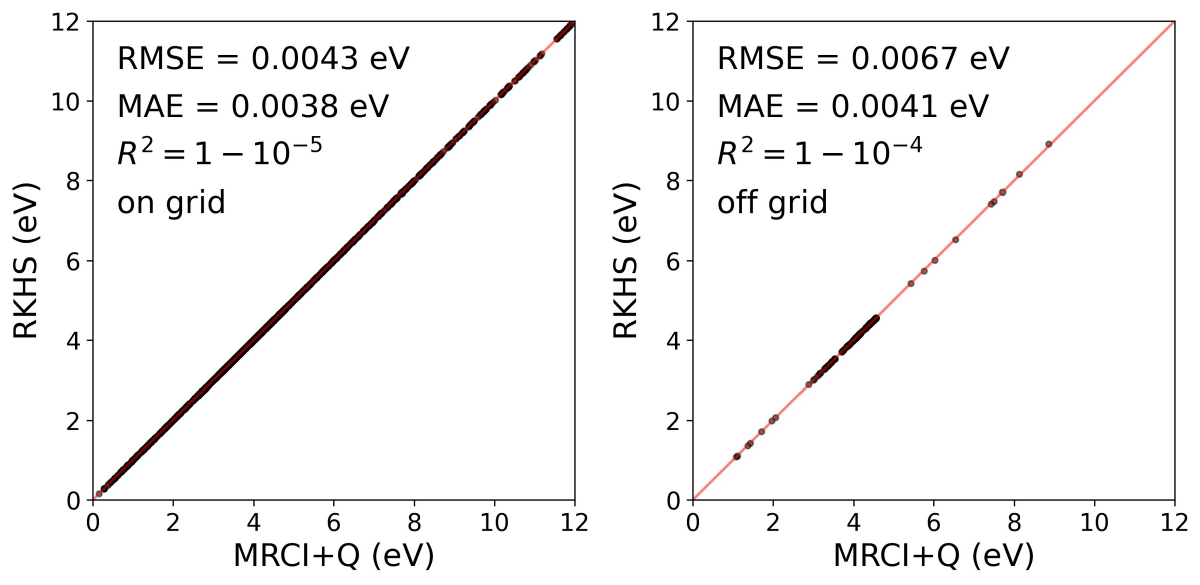}
    \caption{Correlation plots for the on and off-grid structures
      between MRCI+Q and RKHS energies at $1 ^2{\rm A}''$ ground
      state, together with the corresponding RMSE, MAE, and $R^2$
      values. The small RMSE and MAE values, along with an $R^2$ value
      close to 1, indicate that the RKHS PES performs well.}
    \label{fig:ocsp12}
\end{figure}

Next, the RKHS representations of the eight lowest electronic states
are discussed. Their performance in terms of MAE (Mean Absolute
Error), RMSE (Root Mean Square Error), and $R^2$ (Coefficient of
Determination), evaluated on both, on-grid and off-grid data, is shown
in Figures~\ref{fig:ocsp12}, and \ref{sifig:ocsp11} to
\ref{sifig:ocsp32}. Table \ref{sitab:RMSE_MAE_R2} provides numerical
values for the statistical measures and reports the $R-$positions and
relative energies of the minima together with the transition state
energy separating the two minima. For the $1 ^2{\rm A}''$ ground
state, the on- and off-grid performance is reported in Figure
\ref{fig:ocsp12}.  The MAE$(E)$ and RMSE$(E)$ for on-grid points are
0.0038 eV and 0.0043 eV over a range of 12 eV with a correlation
coefficient of $R^2 = 1 - 10^{-5}$. As a validation, the performance
of this RKHS was assessed on the off-grid data set with MAE$(E)$ and
RMSE$(E)$ of 0.0041 eV and 0.0067 eV and $R^2 = 1 - 10^{-4}$. This
deviates only little from the on-grid performance and establishes the
high accuracy of the representation. Compared with other, similar
systems such as CO$_2$, N$_3$, O$_3$, or NO$_2$, this performance is
on par.\cite{MM.co2:2021,MM.n3:2024,MM.o3:2025,MM.no2:2020}\\

\begin{table}[H]
\centering
\begin{tabular}{c|c|c|c|c|c|c||c|c|c|c}
\hline
\cline{2-7} \cline{8-11}
 & \multicolumn{2}{c|}{RMSE} & \multicolumn{2}{c|}{MAE} & \multicolumn{2}{c||}{$R^2$} & \multicolumn{4}{c}{Properties}\\
\hline
 & on & off & on & off & on & off & $R_{\rm min}^{\rm I}$ & $R_{\rm min}^{\rm II}$ & $\Delta E_{\rm I/II}$ & $\Delta E_{\rm I/TS}$ \\
\hline
$1 ^2{\rm A}'$ & 8.5 & 10.2 & 6.8 & 5.2 & $1-10^{-5}$ & $1-10^{-4}$ & 4.32 & 4.38 & 2.96 & 3.57 \\
$2 ^2{\rm A}'$ & 4.3 & 17.0 & 3.8 & 6.7 & $1-10^{-5}$ & $1-10^{-4}$ & 6.14 & 6.09 & 0.32 & 0.47 \\
$3 ^2{\rm A}'$ & 4.5 & 29.0 & 3.9 & 10.3 & $1-10^{-5}$ & $1-10^{-3}$ & 6.02 & 5.99 & 0.38 & 0.51 \\
$4 ^2{\rm A}'$ & 4.4 & 32.0 & 3.8 & 12.2 & $1-10^{-6}$ & $1-10^{-3}$ & 7.04 & 6.31 & 0.19 & 0.33 \\
$1 ^2{\rm A}''$ & 4.3 & 6.7 & 3.8 & 4.1 & $1-10^{-5}$ & $1-10^{-4}$ & 4.32 & 4.36 & 2.95 & 3.36 \\
$2 ^2{\rm A}''$ & 4.3 & 14.3 & 3.8 & 5.6 & $1-10^{-6}$ & $1-10^{-4}$ & 5.99 & 5.99 & 0.52 & 0.66 \\
$3 ^2{\rm A}''$ & 4.3 & 20.2 & 3.8 & 6.8 & $1-10^{-6}$ & $1-10^{-4}$ & 6.15 & 6.09 & 0.31 & 0.47 \\
$1 ^4{\rm A}''$ & 4.3 & 7.0 & 3.8 & 4.4 & $1-10^{-6}$ & $1-10^{-5}$ & 6.36 & 6.20 & 0.22 & 0.37 \\
\hline
\end{tabular}
\caption{RMSE (meV), MAE (meV), and $R^2$ for on- and off-grid points
  for each PES. Also shown are the positions of the primary (I) and
  secondary (II) minima $R$ (a$_0$), the energy difference between the
  two minima $\Delta E_{\rm I/II}$ (eV), and the height of the TS
  originating from minimum I $\Delta E_{\rm I/TS}$ (eV).}
\label{sitab:RMSE_MAE_R2}
\end{table}

\begin{figure}[H]
    \centering
    \includegraphics[width=1.0\linewidth]{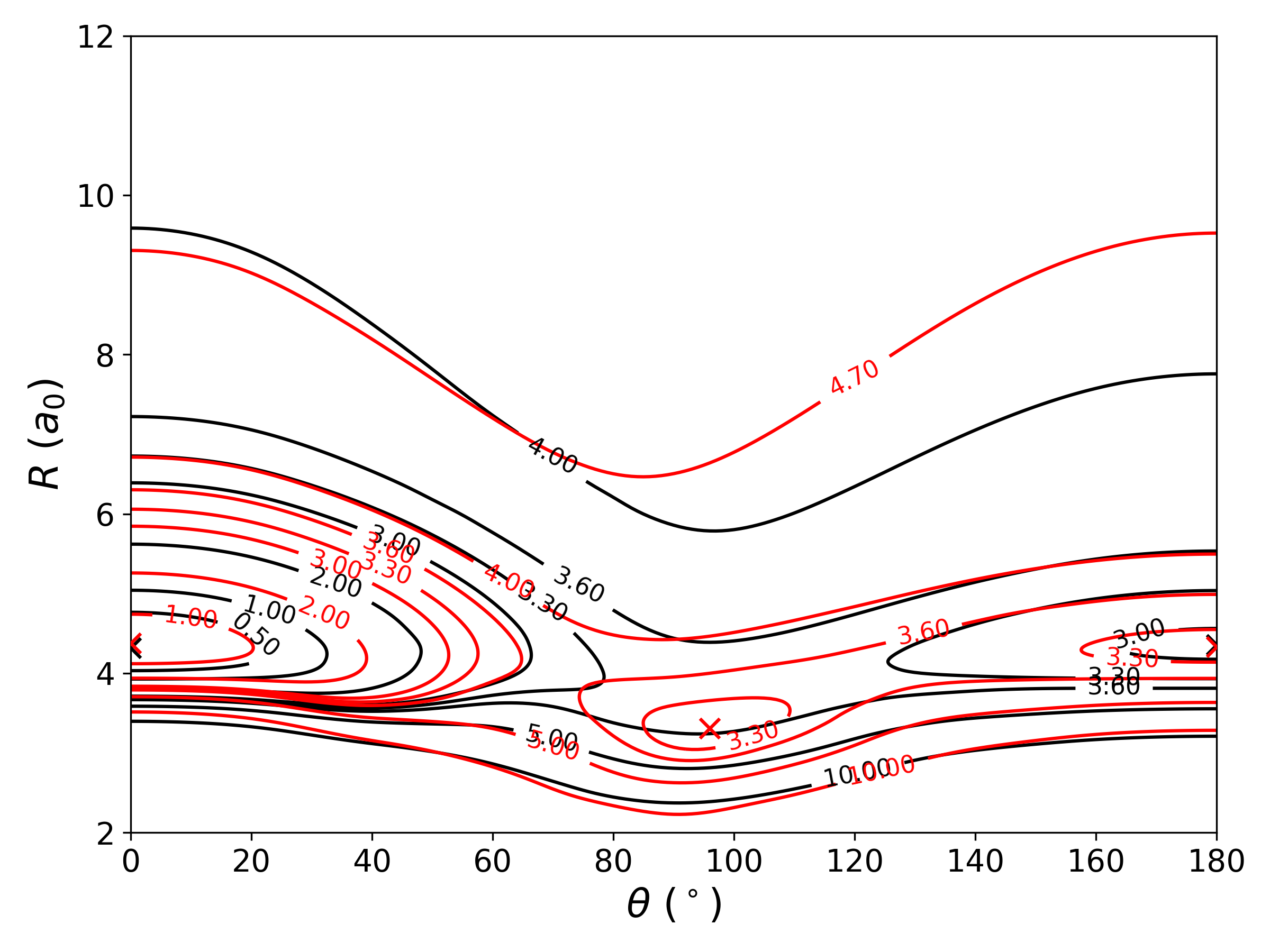}
    \caption{Contour plot of the $1 ^2{\rm A}''$ ground state PES at
      $r = 2.15~a_0$ (equilibrium CO bond, black) and $r = 2.38~a_0$
      (stretched CO bond, red). The numbers on the contour lines
      indicate the energies in eV, and the crosses mark the
      minima. For $r = 2.15~a_0$, the left cross corresponds to the
      global minimum and the right cross to the second minimum. When
      the CO bond is stretched, the S$^+$ atom can insert between the
      C and O atoms, creating a third minimum in addition to the two
      minima at $\theta = 0^\circ$ and $180^\circ$.}
    \label{fig:contour_r}
\end{figure}

\noindent
The $1 ^2{\rm A}''$ ground-state PES of [OC-S]$^+$ exhibits a first
linear minimum energy geometry structure with $r = 2.15~a_0$, $\theta
= 0^\circ$, and $R_{\rm min}^{\rm I} = 4.32$ a$_0$. This minimum
energy structure represents the global minimum and is the zero of
energy for all subsequent analyses and comparisons, see Figure
\ref{fig:contour_r}. A second linear minimum with same $r = 2.15~a_0$
is located at $\theta = 180^\circ$ and $R_{\rm min}^{\rm II} =
4.36$~a$_0$, with an energy $\Delta E_{\rm I/II} = 2.95$ eV above the
global minimum. This structure corresponds to the [S-OC]$^+$
configuration in which the oxygen atom is in the center of the
triatom. The transition state (TS) between these two minima is $\Delta
E_{\rm I/TS} = 3.36$ eV above the global minimum, located at $\theta =
91^\circ$ and $R = 3.67$~a$_0$.\\

\noindent
As the CO bond length increases, a third minimum emerges, and the
features of the PES become more pronounced. For sufficiently stretched
CO bond lengths, this minimum becomes clearly identifiable. When the
C--O bond is stretched to $r = 2.38~a_0$, corresponding to the outer
turning point of the longest CO stretching motion considered in this
work ($v=2$), the ``minimum I'' shifts to $\theta = 0^\circ$, $R_{\rm
  min}^{\rm I} = 4.38$~a$_0$ and its energy rises to 0.63 eV above the
global minimum. Concomitantly, a new local minimum appears at $\theta
= 96^\circ$, $R_{\rm min}^{\rm III} = 3.31$~a$_0$ with energy 3.13 eV
above the global minimum. This ``T-shaped'' structure corresponds to
incipient S$^+$ insertion into the CO bond. ``Minimum II'' appears now
at $\theta = 180^\circ$, $R_{\rm min}^{\rm II} = 4.33$~a$_0$, with an
energy of 3.23 eV above the global minimum. This analysis shows that
stretching the CO bond increases the anisotropy of the PES, introduces
a third minimum, and raises the energy of the linear minima.\\

\noindent
To further validate the RKHS-representation of the $1 ^2{\rm A}''$
ground state PES, a one-dimensional cut $V(R; r = 2.15~{\rm a}_0,
\theta = 1^\circ)$ was computed using the RKHS-fitted PES and compared
with MRCI+Q/aVTZ calculations, see brown trace and open brown circles
in Figure \ref{sifig:pes_lines_and_circles}. The overall satisfactory
agreement between the RKHS curve and the \textit{ab initio} points
demonstrates that the RKHS-PES performs reliably for on- and off-grid
points and for particularly relevant (off-grid) cuts such as the
linear configuration which was not part of the 3d-grid. The RMSD is
0.072 eV for the ground state, while all other electronic states
considered here confirm this finding with RMSD values ranging from
0.075 to 0.557 eV.\\

\begin{figure}[H]
    \centering
    \includegraphics[width=1.0\linewidth]{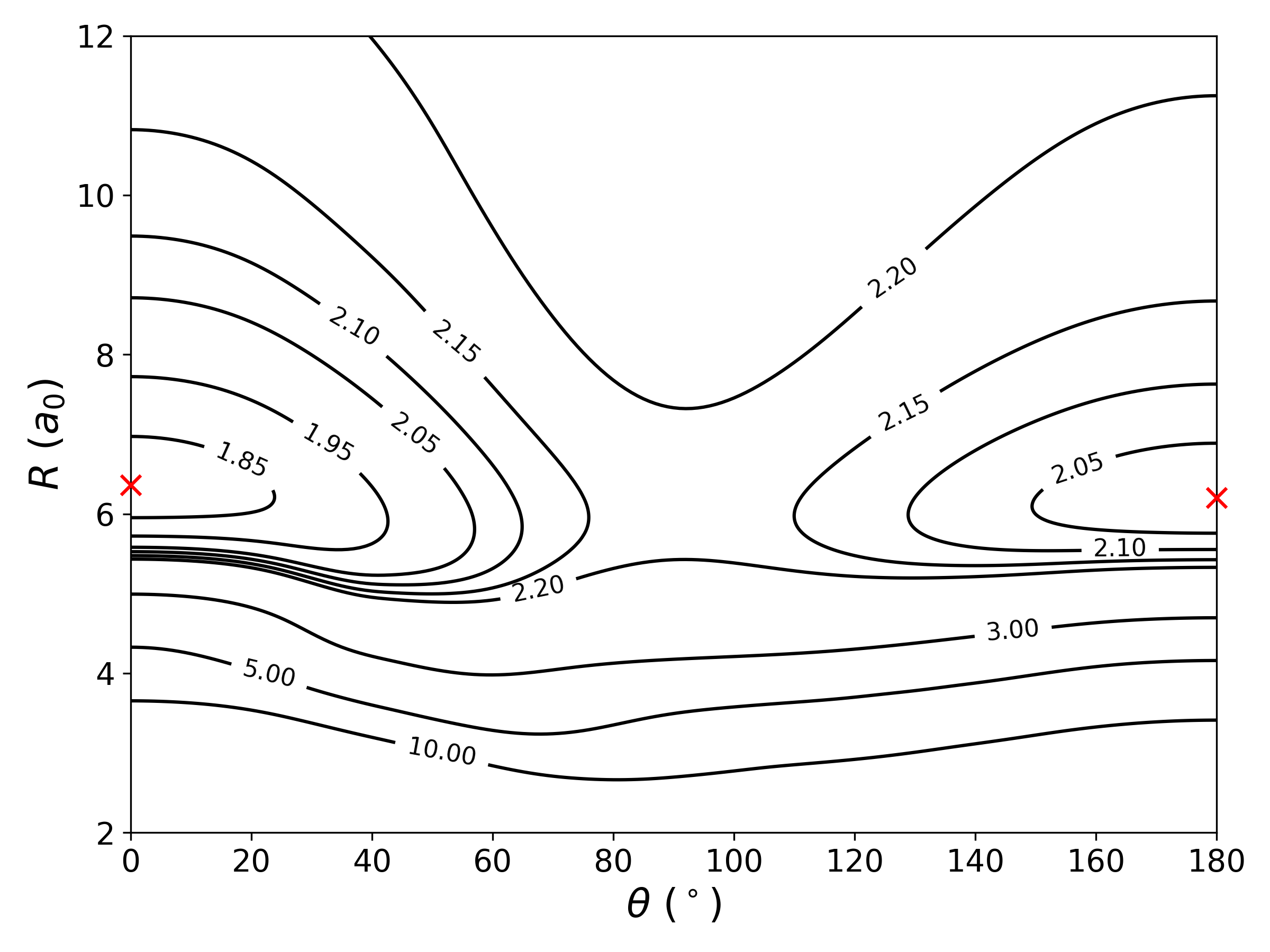}
    \caption{Contour plot of the $1 ^4 {\rm A} ''$ state PES for $r =
      2.15~a_0$. The numbers on the contour lines indicate the
      energies in eV, and the crosses mark the minima. Compared to the
      $1 ^2{\rm A}''$ state, the minima of the $1 ^4 {\rm A} ''$ state
      are noticeably shallower. At long range, this state corresponds
      to the CO($\tilde{\mathrm{X}}~^1\Sigma^+$)+S$^+$($^4$S$_u$)
      dissociation limit.}
    \label{fig:contour_quartet}
\end{figure}

\noindent
Next, the topology of the $1 ^4{\rm A}''$ state correlating with the
lowest energy state for $R \rightarrow \infty$ is discussed. This
RKHS-PES exhibits an equilibrium geometry at $r = 2.15~a_0$ and
$\theta = 0^\circ$, with $R_{\rm min}^{\rm I} = 6.36$~a$_0$ and a
global minimum which lies 1.80 eV above the minimum of the $1 ^2{\rm
  A}''$ ground state. This PES is considerably flatter, with a
dissociation energy of only 0.43 eV, compared with 4.10 eV for the
ground state PES. Minimum II is located at $r = 2.15~a_0$ with $\theta
= 180^\circ$ and $R_{\rm min}^{\rm II} = 6.20$~a$_0$, and lies 0.22 eV
above minimum I, see Figure \ref{fig:contour_quartet}. These results
are in good agreement with the \textit{ab initio} data shown in Figure
\ref{fig:pes-1d}.\\

\begin{table} [H]
    \centering
    \begin{tabular}{c|c|c|c|c}
    \hline
    & CASPT2/ANO-L\cite{chen:2006} & MRCI/VTZ\cite{hirst:2006} & MRCI+Q/aVTZ & Expt.\cite{eland:1973,wang88a}\\
    \hline
    OCS$^+$($\tilde{\mathrm{X}}~^2\Pi$) & 0.00 & 0.00 & 0.00 & 0.00\cite{eland:1973},0.00\cite{wang88a} \\
    OCS$^+$($\tilde{\mathrm{A}}~^2\Pi$) & 3.79 & 4.05 & 3.88 & 3.89\cite{eland:1973},3.89\cite{wang88a} \\
    OCS$^+$($\tilde{\mathrm{B}}~^2\Sigma^+$) & 4.91 & 5.42 & 4.61 & 4.85\cite{eland:1973},4.86\cite{wang88a} \\
    CO($\tilde{\mathrm{X}}~^1\Sigma^+$)+S$^+$($^4$S$_u$) & 2.01 &  & 2.14 & 2.31\cite{eland:1973} \\
    CO($\tilde{\mathrm{X}}~^1\Sigma^+$)+S$^+$($^2$D$_u$) & 3.92 &  & 4.04 & 4.15\cite{eland:1973} \\
    CO($\tilde{\mathrm{X}}~^1\Sigma^+$)+S$^+$($^2$P$_u$) & 5.24 &  & 5.19 & 5.36\cite{eland:1973} \\
    \hline
    \end{tabular}
    \caption{Excitation energies for the first two electronically
      excited-states of OCS$^+$ and the three lowest dissociation
      asymptotes. Energies are reported relative to the ground state
      of OCS$^+$, in eV. Note that the MRCI+Q/aVTZ values reported in
      this table were obtained from the 3D PESs and therefore cannot
      be directly identified from the 1d-cuts shown in Figure
      \ref{fig:pes-1d}.}
    \label{tab:OCSenergies}
\end{table}

\noindent
The adiabatic excitation energies of the reactants and asymptotic
products relative to the ground-state reactant
OCS$^+$($\tilde{\mathrm{X}}~^2\Pi$) are summarized in
Table~\ref{tab:OCSenergies}. The MRCI+Q/aVTZ method yields
dissociation energies for the CO+S$^+$ products in different
electronic states in close agreement with experiment,\cite{eland:1973}
with largest deviations of about 0.17 eV. This represents an
improvement over the CASPT2/ANO-L results, which show a largest
deviations of approximately 0.30 eV.\cite{chen:2006} Specifically,
CASPT2 systematically underestimates the asymptotic energies,
predicting 2.01, 3.92, and 5.24 eV for
CO($\tilde{\mathrm{X}}~^1\Sigma^+$)+S$^+$($^4$S$_u$),
CO($\tilde{\mathrm{X}}~^1\Sigma^+$)+S$^+$($^2$D$_u$), and
CO($\tilde{\mathrm{X}}~^1\Sigma^+$)+S$^+$($^2$P$u$), respectively,
compared to the experimental values of 2.31, 4.15, and 5.36 eV. In
contrast, MRCI+Q/aVTZ provides improved agreement for the dissociation
limits, yielding 2.14, 4.04, and 5.19 eV. \\

\subsection{Crossing Manifolds of the PESs}
\begin{figure}[H]
    \centering \includegraphics[width=0.8\linewidth]{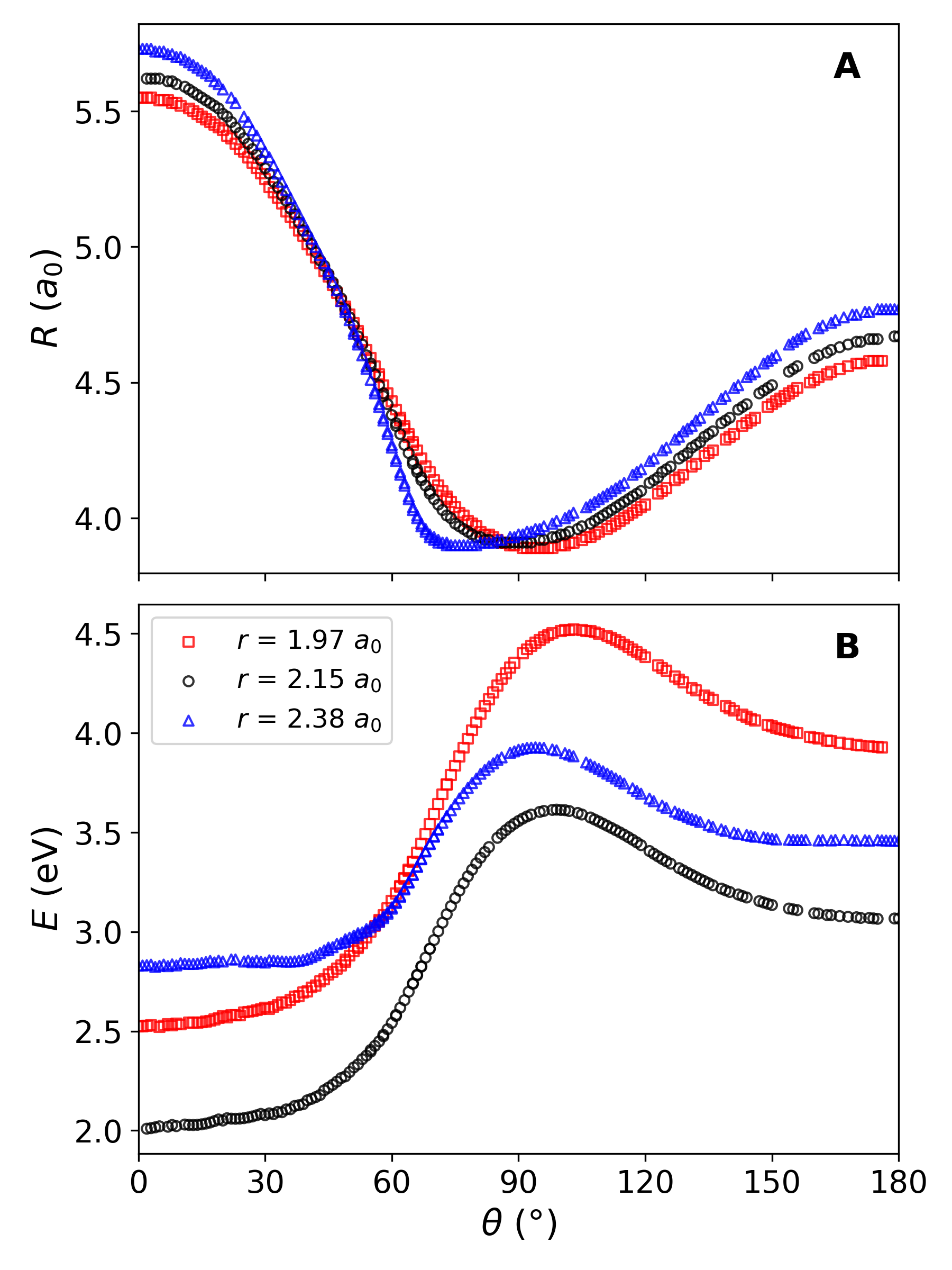}
    \caption{The state-crossing region between the $1 ^2{\rm A}'$ and
      $1 ^4{\rm A}''$ states at three fixed CO bond length of $r =
      [1.97, 2.15, 2.38]$ a$_0$. Here, $r = 2.15~a_0$ corresponds to
      the equilibrium structure and the two other values are the
      classical turning points for the $v=0$ vibrational ground
      state. Each point in the figure represents a geometry for which
      $\Delta E_{I/II} < 0.01$ eV, where I and II are the two PESs
      considered. Panel A: Crossing point along $R$ as a function of
      $\theta$. Panel B: Crossing energies depending on $\theta$ for
      the three CO-separations considered in panel A.}
    \label{fig:cross1}
\end{figure}

As can be already seen from Figure \ref{fig:pes-1d}A, in the region $R
\in [4,5]$ a$_0$, the PESs feature multiple curve crossings. In
addition, the $1 ^2{\rm A}'$ and $1 ^2{\rm A}''$ states, which are
very close in energy, both cross with the $1 ^4{\rm A}''$ state.  The
location of the crossing points also depends on the angle $\theta$, as
seen in Figure \ref{fig:pes-1d}B. Hence, these crossing manifolds,
which in general are 2-dimensional, are considered next. Figure
\ref{fig:cross1} reports the $R-$values for the $1 ^2{\rm A}'$ / $1
^4{\rm A}''$ crossing for different values of the CO bond length,
including the classical turning points (red and blue) for the $v_{\rm
  CO} = 0$ ground state and the equilibrium value of the CO separation
(black). The crossing seams all feature a pronounced angular
anisotropy which also mirrors the anisotropies of the two
PESs. Contracting the CO bond away from the minimum energy structure
changes the crossing seam insignificantly, whereas extending the CO
separation (blue curve) leads to a more pronounced change around the
T-shaped configuration.\\

\noindent
The state-crossing region between other states are shown in Figure
\ref{sifig:manifold_1} to Figure \ref{sifig:manifold_3}. The behavior
is consistent with the one-dimensional projections shown in Figure
\ref{fig:pes-1d}.\\

\subsection{Exploratory QCT Simulations}
To further validate the reactive 3d-PESs exploratory QCT simulations
were carried out. For this, 10000 trajectories were run on each of the
$1 ^2{\rm A}''$ (ground state of the OCS$^+$ complex) and $1 ^4{\rm
  A}''$ (ground state in the asymptotic limit) PESs.\\

\begin{figure}[H]
    \centering
    \includegraphics[width=1.0\linewidth]{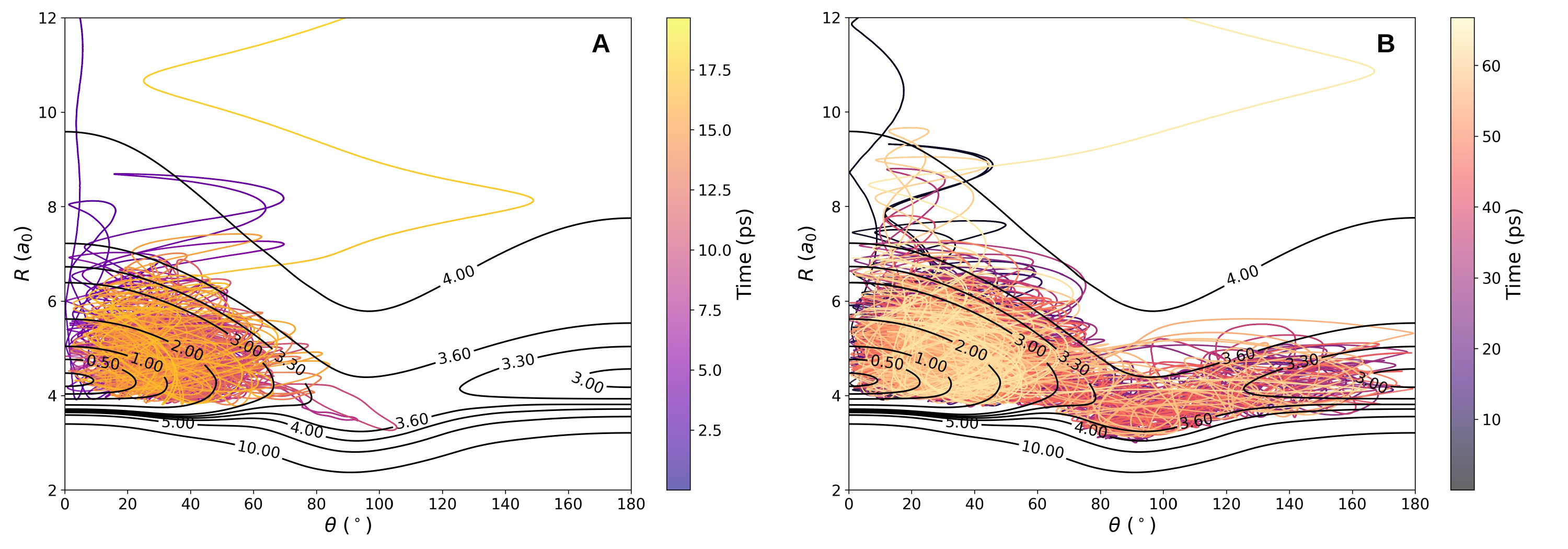}
    \caption{Two representative trajectories mapped on the $1 ^2{\rm
        A}''$ PES. Panel A illustrates a trajectory leading to the
      formation of the global minimum, linear OCS$^+$. Panel B shows a
      different trajectory in which, after OCS$^+$ is formed, the
      system also passes through the secondary minimum, linear
      S$^+$OC. In a related system - CO$_2$ - the existence of a
      secondary local OOC minimum was found from calculations at the
      MRCI+Q and CCSD(T) levels of theory.\cite{MM.co2:2021}.}
    \label{fig:cont_traj}
\end{figure}

\noindent
Typical trajectories mapped onto the PESs are shown in Figure
\ref{fig:cont_traj}. Panel A illustrates a trajectory leading to the
formation of the global minimum, linear OCS$^+$. Initially, CO and
S$^+$ are placed far apart (30 $a_0$); they collide to form linear
OCS$^+$, vibrate around the minimum for approximately 20 ps, and
eventually dissociate back into CO and S$^+$. Panel B shows a
different trajectory in which, after linear OCS$^+$ is formed, the
system also passes through the secondary minimum, linear S$^+$OC,
before dissociating after 60 ps.\\

\noindent
To further validate the PESs, particular characteristics and
observables were computed from the QCT simulations. First, the
conservation of total energy was evaluated, see Figure
\ref{sifig:cons_hist}. For each trajectory the fluctuation around the
mean for the total energy was determined and the typical deviations
are $\sim 5$ cm$^{-1}$ or less. No energy drift was observed in all
runs. Next, the probability for reaction depending on the impact
parameter $b$ was computed. The probability for reaction $P(b)$ - also
known as the opacity function - for the reactive trajectories was also
determined, see Figure \ref{sifig:opacity_pb}. On the time scale of
the present simulations (1 ns) all the [OCS]$^+$ complexes
decay. Interestingly, $P(b)$ has a nontrivial structure which may be
worth analyzing. As can be seen, the interaction potential is rather
long-ranged with $b_{\rm max} \sim 20$ a$_0$ at $E_{\rm coll} = 0.05$
eV. The rotational state undergoes significant change during the
dynamics: while the initial condition is set to the ground state
($j=0$), the final rotational states span a broad range from 0 to
50. As shown in Figure \ref{sifig:j_dis}, this is reflected in the
product-state rotational distribution function, $P(j')$. And it is
related to the initial to $\theta$, see Figure
\ref{sifig:j_vs_theta}. Together, these quantities provide a
complementary and balanced description of the underlying dynamical
behavior.\\

\noindent
The CO distance and SOC angle distributions are evaluated by
extracting the geometries from trajectories obtained in 5 independent
simulations and are shown in Figure \ref{sifig:r_co_and_a_soc}. These
distributions provide insight into the configurational space sampled
during the dynamics and help justify the range of geometries included
in the simulations, particularly the extent of CO bond distortion and
the preferred orientation of the approaching S atom.\\

\section{Conclusions}
This work presents a comprehensive study of the reactive PESs of the
OCS$^+$ cation, focusing on S$^+$-loss in the electronic ground state
and seven low-lying electronically excited states. High-level
\textit{ab initio} calculations were performed at the
MRCI+Q/aug-cc-pVTZ level in Jacobi coordinates, providing reference
energies on a dense 3D grid. These data were used to construct smooth,
full-dimensional PESs using RKHS representations, accurately
reproducing dissociation limits to CO + S$^+$ and agreeing closely
with experimental results and previous theoretical studies.\\

\noindent
The PESs reveal rich topological features, including linear and
T-shaped minima at each states, pronounced angular anisotropy, and
extended state-crossing manifolds sensitive to the molecular
geometry. Detailed analyses of the crossings between the $1 ^2{\rm
  A}'$ and $1 ^4{\rm A}''$ states highlight the 2D nature and
angle-dependence of these manifolds. Exploratory QCT simulations
confirm the numerical stability of the PESs, energy conservation, and
provide insight into reactive dynamics, including rotational
distributions and opacity functions.\\

\noindent
Overall, this work provides the most complete characterization of the
low-lying PESs of OCS$^+$ to date and lays a robust foundation for
future studies of photodissociation and chem-ionization
dynamics.\\

\section*{Supporting Information} 
The supporting information provides additional relevant figures.\\

\section*{Acknowledgment}
The authors gratefully acknowledge financial support from the Swiss
National Science Foundation through grants $200020\_219779$ (MM),
$200021\_215088$ (MM), the NCCR-MUST (MM), and the University of Basel
(MM). This article is also based upon work within COST Action COSY
CA21101, supported by COST (European Cooperation in Science and
Technology) (to MM).\\

\section*{Data Availability}
The codes and data for the present study are available from
\url{https://github.com/MMunibas/OCS} upon publication.

\clearpage

\renewcommand{\thetable}{S\arabic{table}}
\renewcommand{\thefigure}{S\arabic{figure}}
\renewcommand{\thesection}{S\arabic{section}}
\renewcommand{\d}{\text{d}}
\setcounter{figure}{0}  
\setcounter{section}{0}  
\setcounter{table}{0}

\newpage

\noindent
\LARGE{\bf SUPPORTING INFORMATION:\\ Full-Dimensional Reactive
  Potential Energy Surfaces for OCS$^+$ $\rightarrow$ CO+S$^+$
  Dissociation: Ground and Excited States}

\newpage

\begin{figure}[h!]
    \centering \includegraphics[width=0.8\linewidth]{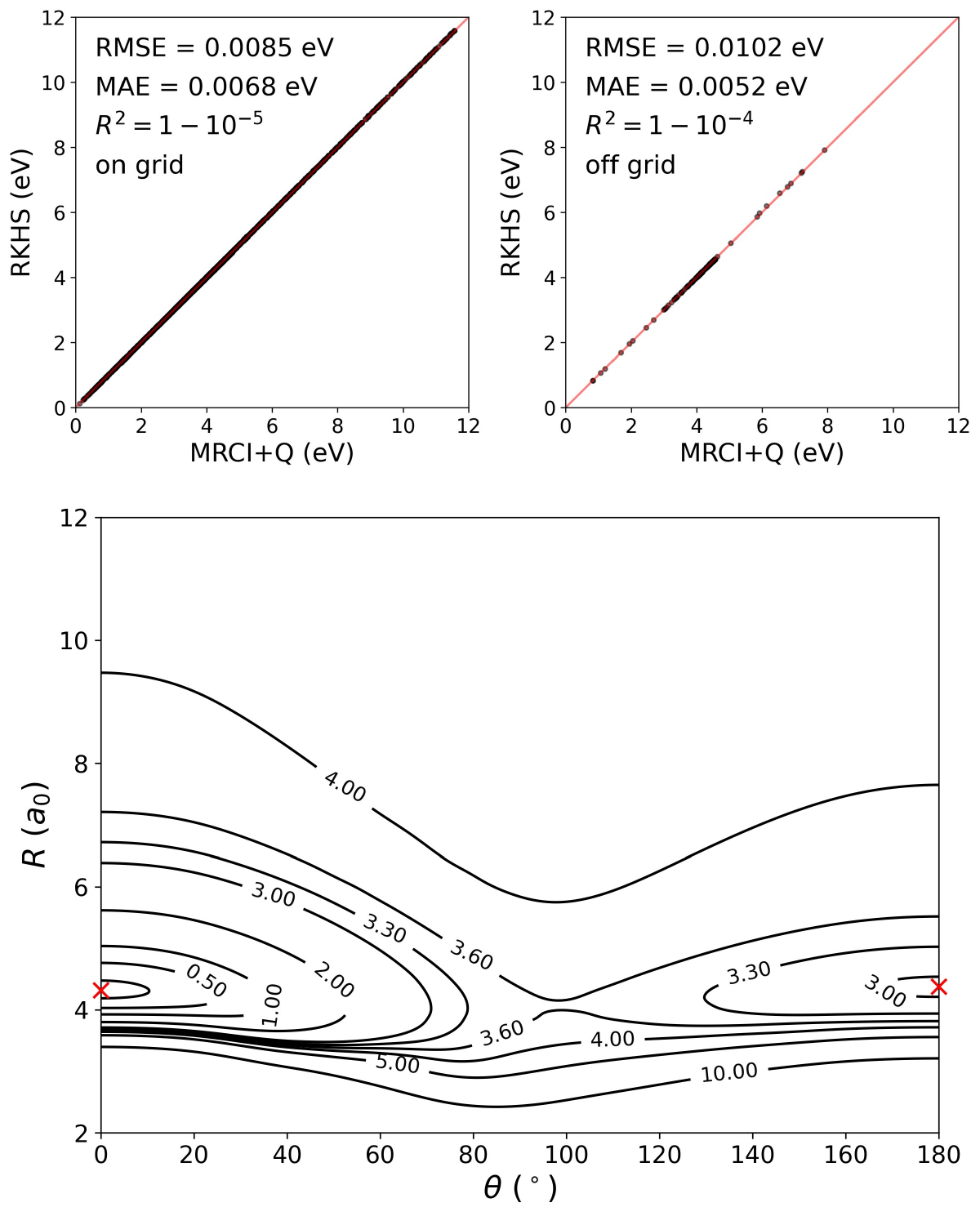}
    \caption{On- and off-grid performance of the RKHS and topography
      for the $1 ^2{\rm A}'$ state. Correlation plots comparing MRCI+Q
      and RKHS energies for on-grid and off-grid structures are shown
      at the top, together with the corresponding RMSE, MAE, and $R^2$
      values. The contour plot at $r = 2.15~a_0$ is displayed at the
      bottom. The numbers on the contour lines indicate energies in
      eV, and the red crosses mark the minima. The two minima are
      identical to those of the $1 ^2{\rm A}''$ state, as
      expected. For the linear configuration, the $1 ^2{\rm A}'$ and
      $1 ^2{\rm A}''$ states are degenerate and correspond to the same
      electronic state.}
    \label{sifig:ocsp11}
\end{figure}

\begin{figure}[h!]
    \centering \includegraphics[width=0.8\linewidth]{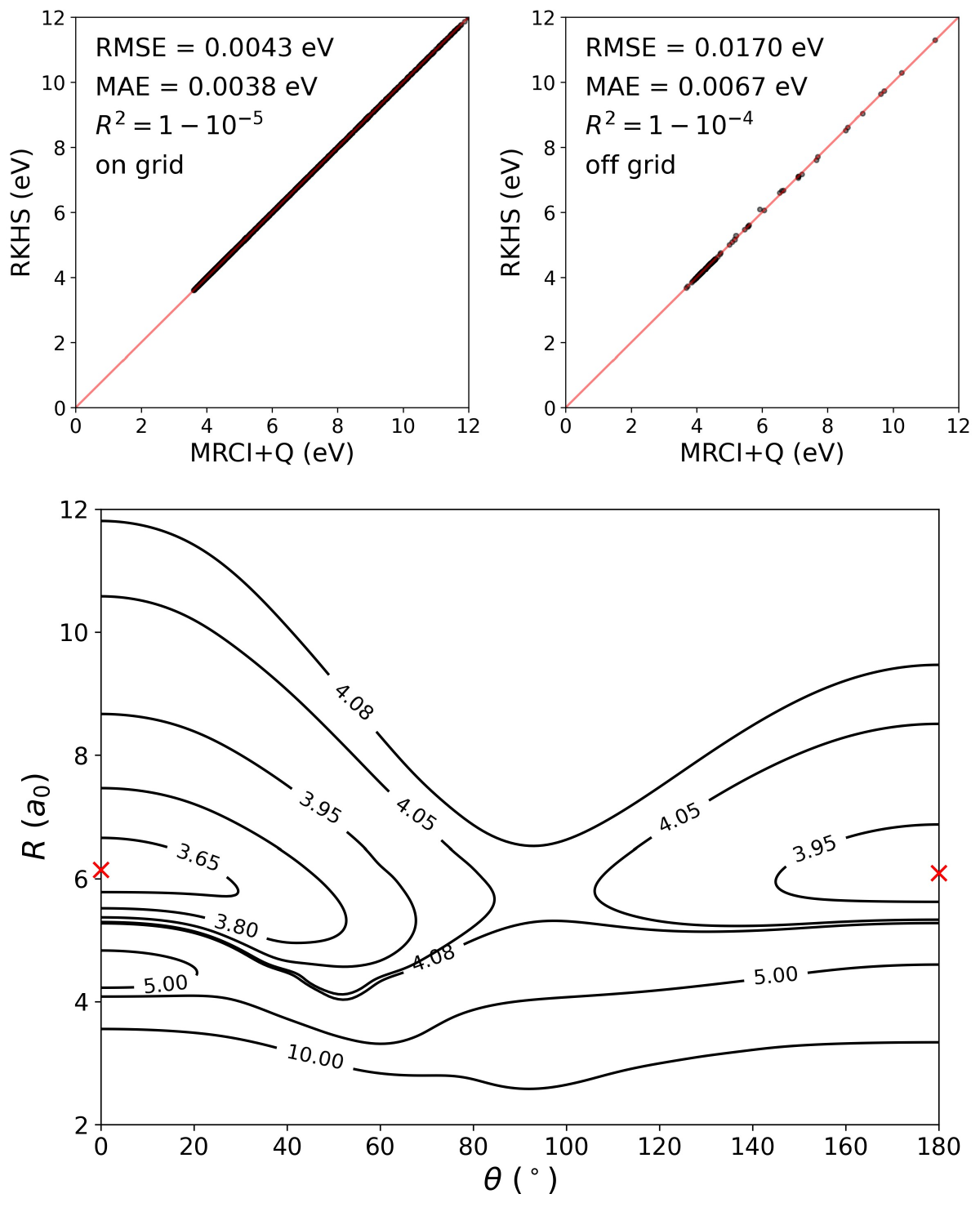}
    \caption{On- and off-grid performance of the RKHS and topography
      for the $2 ^2{\rm A}'$ state. Correlation plots comparing MRCI+Q
      and RKHS energies for on-grid and off-grid structures are shown
      at the top, together with the corresponding RMSE, MAE, and $R^2$
      values. The contour plot at $r = 2.15~a_0$ is displayed at the
      bottom. The numbers on the contour lines indicate energies in
      eV, and the red crosses mark the minima. Compared to the $1
      ^2{\rm A}'$ state, the minima of the $2 ^2{\rm A}'$ state are
      noticeably shallower. At long range, this state becomes
      degenerate with the $1 ^2{\rm A}''$ and $1 ^2{\rm A}'$ states,
      corresponding to the
      CO($\tilde{\mathrm{X}}~^1\Sigma^+$)+S$^+$($^2$D$_u$)
      dissociation limit.}
    \label{sifig:ocsp21}
\end{figure}

\begin{figure}[h!]
    \centering \includegraphics[width=0.8\linewidth]{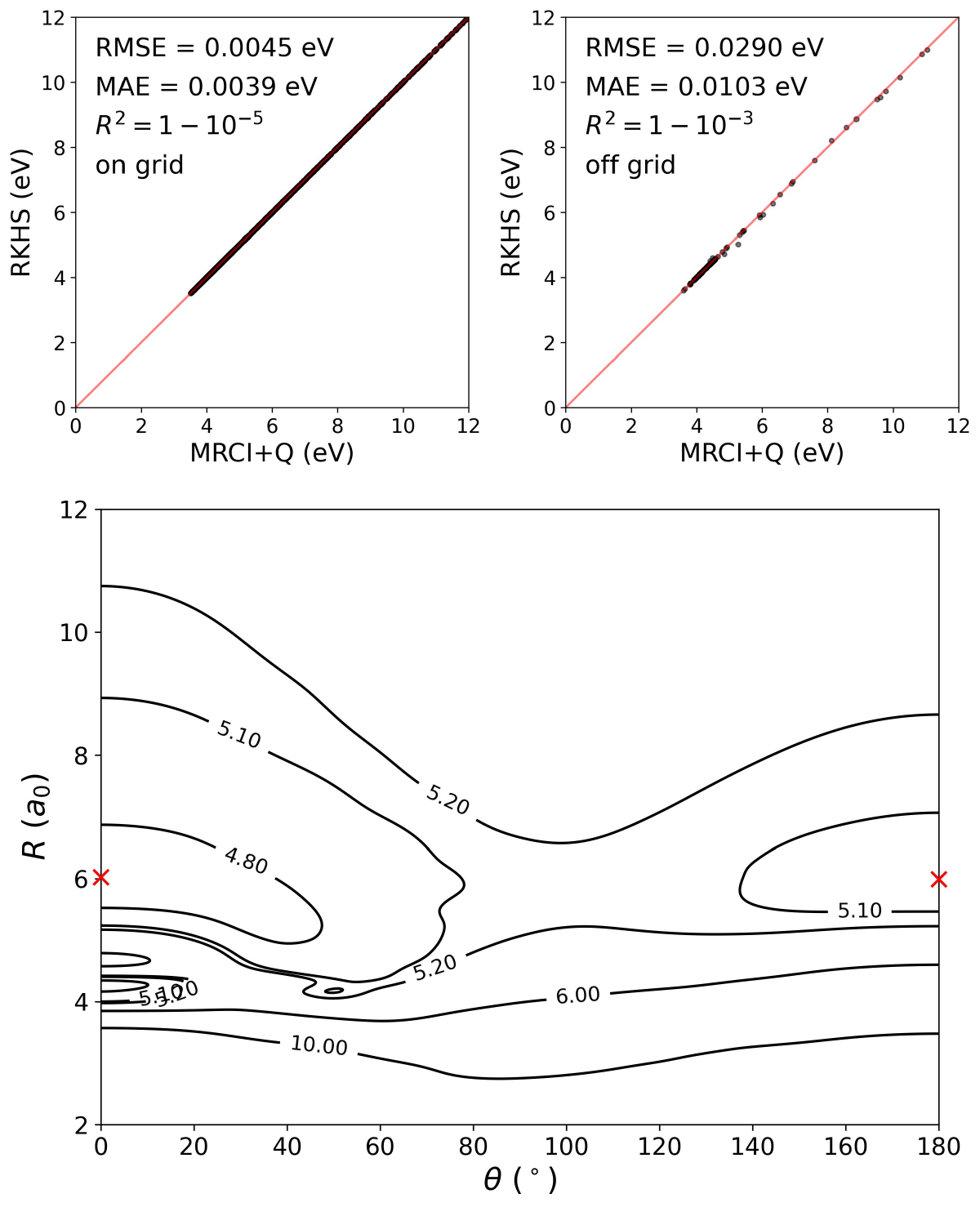}
    \caption{On- and off-grid performance of the RKHS and topography
      for the $3 ^2{\rm A}'$ state. Correlation plots comparing MRCI+Q
      and RKHS energies for on-grid and off-grid structures are shown
      at the top, together with the corresponding RMSE, MAE, and $R^2$
      values. The contour plot at $r = 2.15~a_0$ is displayed at the
      bottom. The numbers on the contour lines indicate energies in
      eV, and the red crosses mark the minima. At long range, this
      state correlates with the
      CO($\tilde{\mathrm{X}}~^1\Sigma^+$)+S$^+$($^2$P$_u$)
      dissociation limit.}
    \label{sifig:ocsp31}
\end{figure}

\begin{figure}[h!]
    \centering \includegraphics[width=0.8\linewidth]{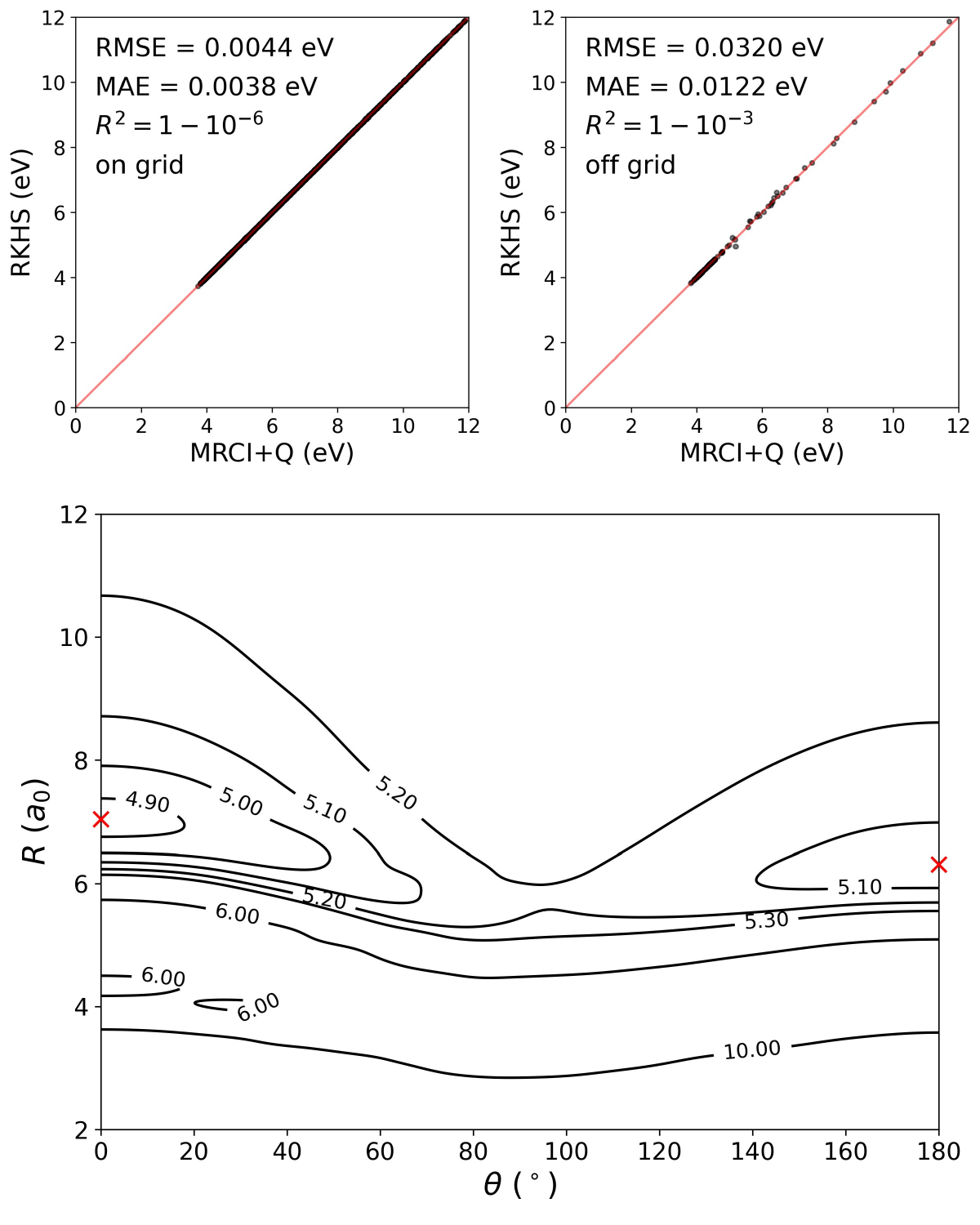}
    \caption{On- and off-grid performance of the RKHS and topography
      for the $4 ^2{\rm A}'$ state. Correlation plots comparing MRCI+Q
      and RKHS energies for on-grid and off-grid structures are shown
      at the top, together with the corresponding RMSE, MAE, and $R^2$
      values. The contour plot at $r = 2.15~a_0$ is displayed at the
      bottom. The numbers on the contour lines indicate energies in
      eV, and the red crosses mark the minima. At long range, this
      state becomes degenerate with the $3 ^2{\rm A}'$ state,
      corresponding to the
      CO($\tilde{\mathrm{X}}~^1\Sigma^+$)+S$^+$($^2$P$_u$)
      dissociation limit.}
    \label{sifig:ocsp41}
\end{figure}

\begin{figure}[h!]
    \centering \includegraphics[width=0.8\linewidth]{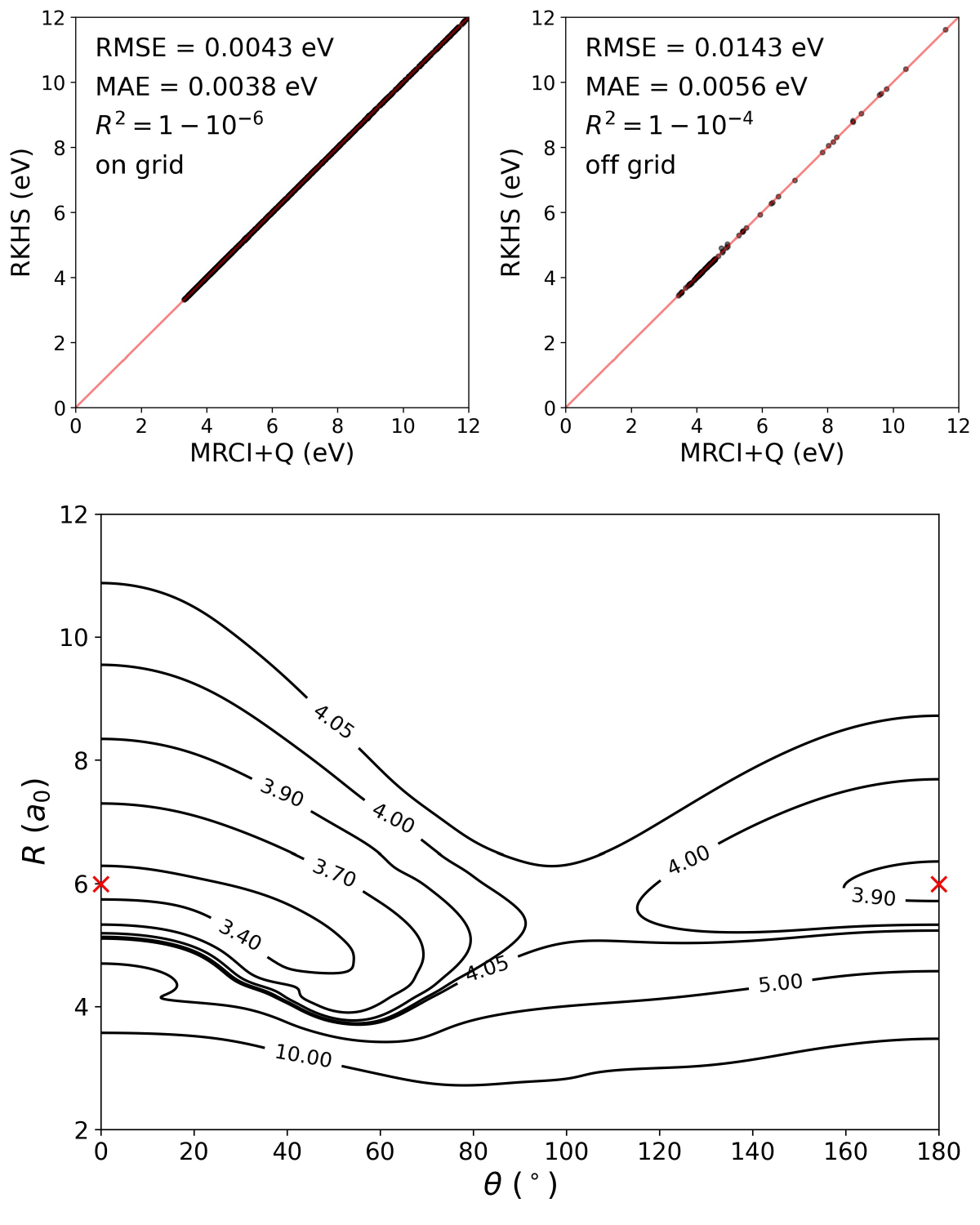}
    \caption{On- and off-grid performance of the RKHS and topography
      for the $2 ^2{\rm A}''$ state. Correlation plots comparing
      MRCI+Q and RKHS energies for on-grid and off-grid structures are
      shown at the top, together with the corresponding RMSE, MAE, and
      $R^2$ values. The contour plot at $r = 2.15~a_0$ is displayed at
      the bottom. The numbers on the contour lines indicate energies
      in eV, and the red crosses mark the minima. At long range, this
      state becomes degenerate with the $1 ^2{\rm A}''$, $1 ^2{\rm
        A}'$, and $2 ^2{\rm A}'$ states, corresponding to the
      CO($\tilde{\mathrm{X}}~^1\Sigma^+$)+S$^+$($^2$D$_u$)
      dissociation limit.}
    \label{sifig:ocsp22}
\end{figure}

\begin{figure}[h!]
    \centering \includegraphics[width=0.8\linewidth]{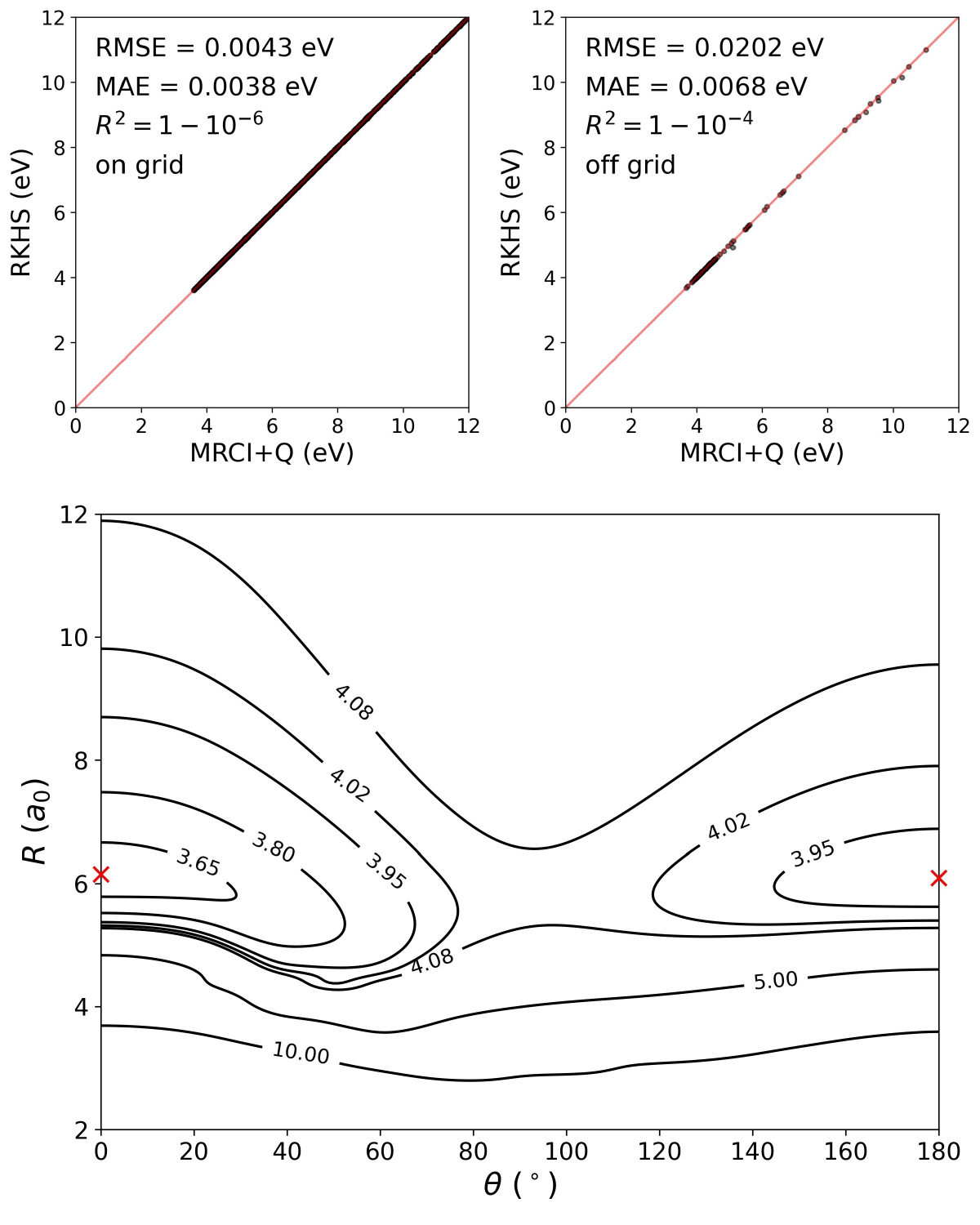}
    \caption{On- and off-grid performance of the RKHS and topography
      for the $3 ^2{\rm A}''$ state. Correlation plots comparing
      MRCI+Q and RKHS energies for on-grid and off-grid structures are
      shown at the top, together with the corresponding RMSE, MAE, and
      $R^2$ values. The contour plot at $r = 2.15~a_0$ is displayed at
      the bottom. The numbers on the contour lines indicate energies
      in eV, and the red crosses mark the minima. At long range, this
      state becomes degenerate with the $1 ^2{\rm A}''$, $1 ^2{\rm
        A}'$, $2 ^2{\rm A}''$, and $2 ^2{\rm A}'$ states,
      corresponding to the
      CO($\tilde{\mathrm{X}}~^1\Sigma^+$)+S$^+$($^2$D$_u$)
      dissociation limit.}
    \label{sifig:ocsp32}
\end{figure}

\begin{figure}[h!]
    \centering
    \includegraphics[width=1.0\linewidth]{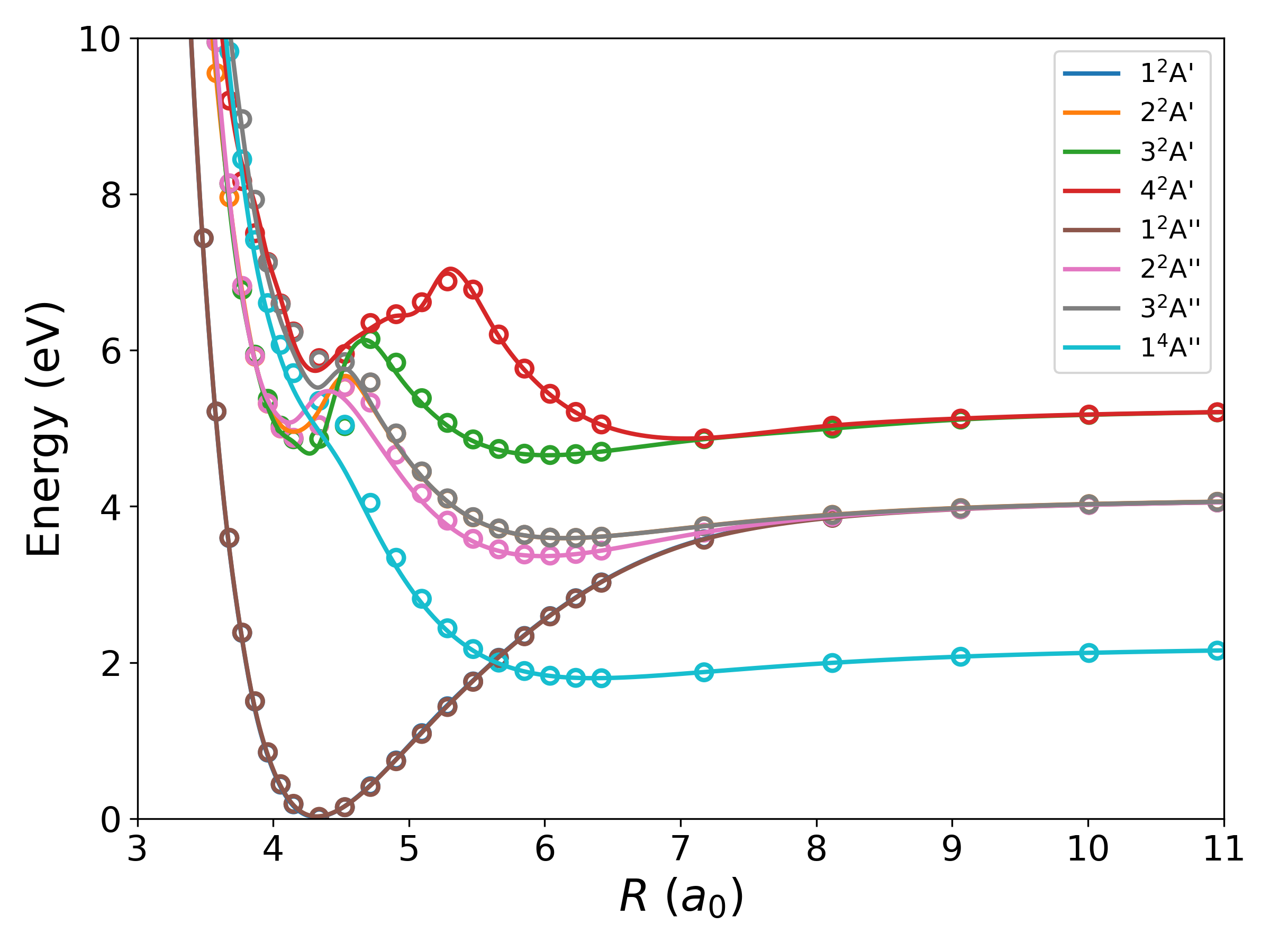}
    \caption{One-dimensional cut of the $V(R; r = 2.15~{\rm a}_0,
      \theta = 1^\circ)$ PES obtained from the RKHS-fitted PES. The
      solid lines represent the RKHS-fitted curves, while the open
      circles correspond to reference on-grid \textit{ab initio}
      energies.}
    \label{sifig:pes_lines_and_circles}
\end{figure}

\begin{figure}[h!]
    \centering \includegraphics[width=1.0\linewidth]{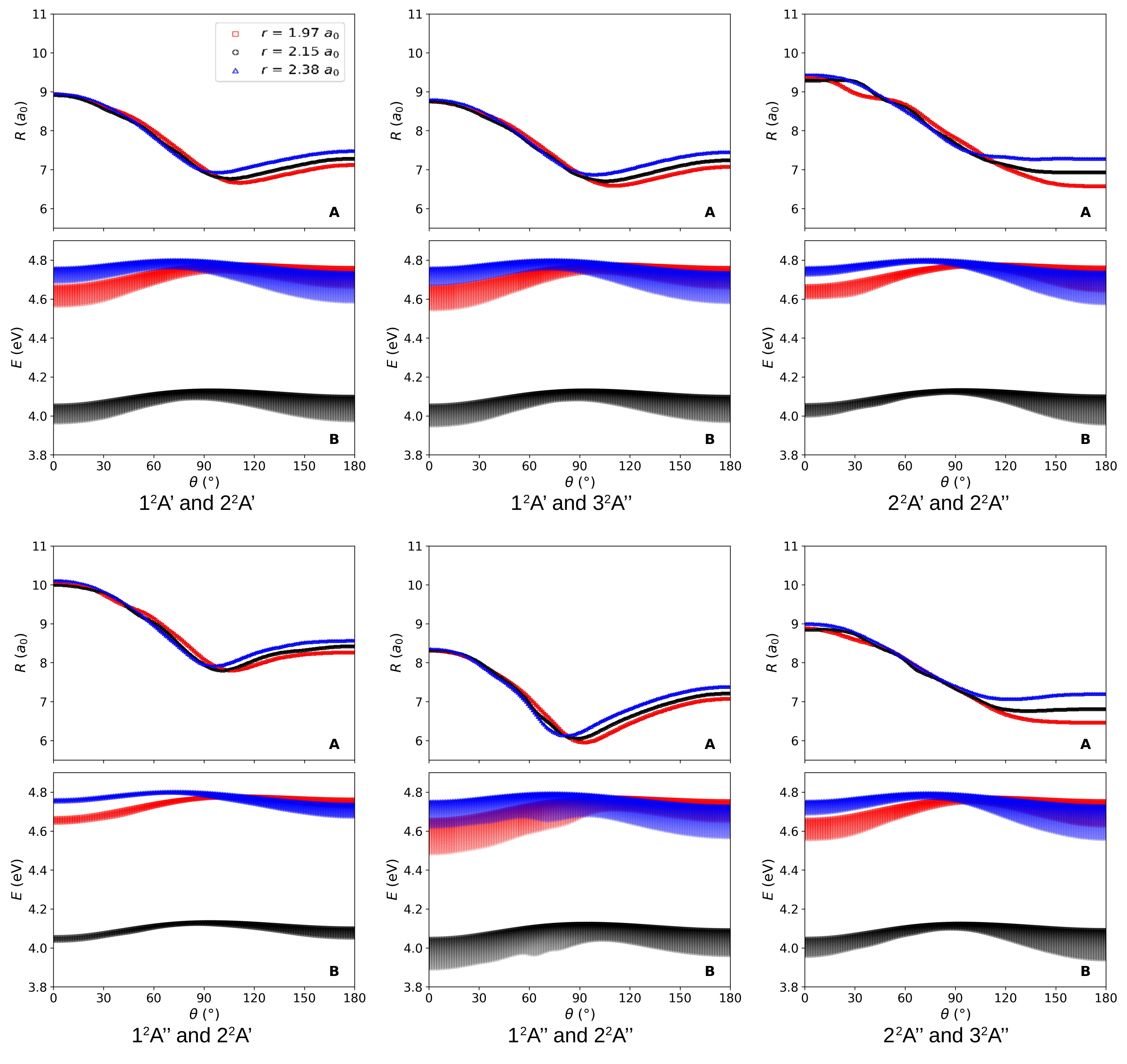}
    \caption{Crossing manifolds between paired states among the $1
      ^2{\rm A}'$, $2 ^2{\rm A}'$, $1 ^2{\rm A}''$, $2 ^2{\rm A}''$,
      and $3 ^2{\rm A}''$ states are shown. Each point in the figure
      corresponds to a geometry at which the two states have the same
      energy, with an energy difference smaller than 0.01 eV. The top
      row reports the lower boundary beyond which the two states
      become degenerate, i.e., for $R$ larger than the boundary value,
      $E_1 \approx E_2$. In the large-$R$ limit, all five states are
      degenerate. However, the onset of degeneracy exhibits a clear
      $\theta$-dependence. These manifolds display very similar
      features, differing only slightly from one another. At the
      equilibrium bond length $r = 2.15~a_0$ (black), the dissociation
      energy is approximately 4.0 eV, whereas at $r = 1.97~a_0$ (red)
      and $r = 2.38~a_0$ (blue), it increases to about 4.5–4.8 eV.}
    \label{sifig:manifold_1}
\end{figure}

\begin{figure}[h!]
    \centering \includegraphics[width=0.5\linewidth]{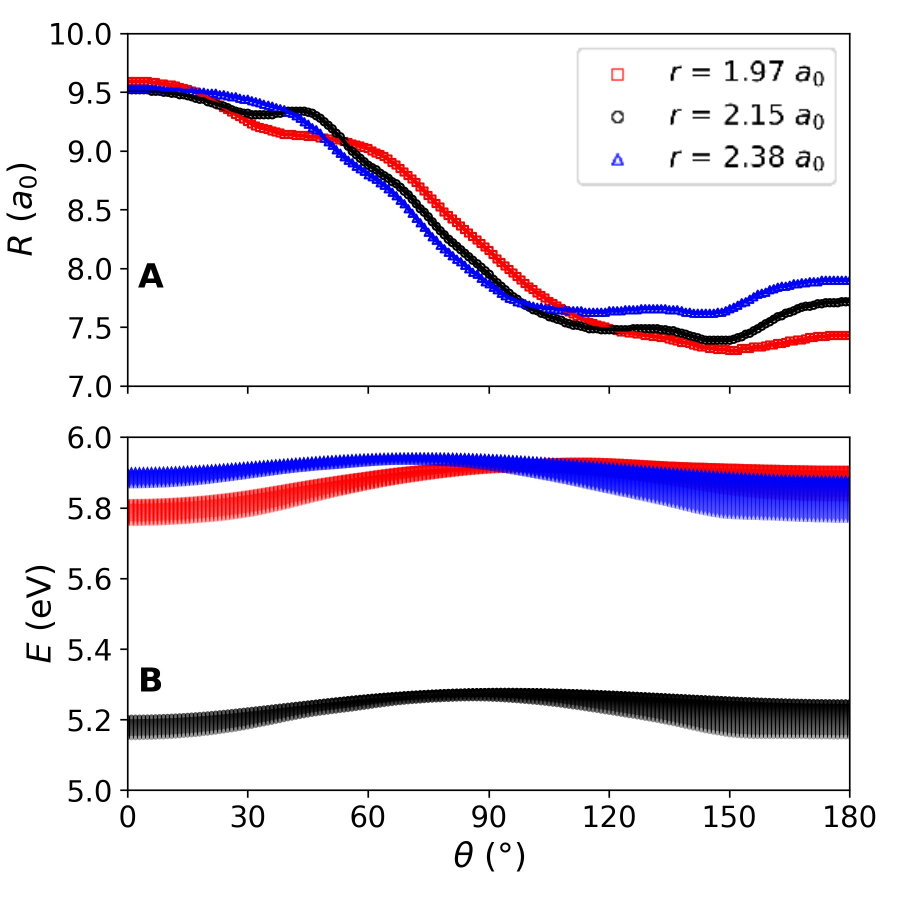}
    \caption{Crossing manifolds between paired states between $3
      ^2{\rm A}'$ and $4 ^2{\rm A}'$ states are shown. Each point in
      the figure corresponds to a geometry at which the two states
      have the same energy, with an energy difference smaller than
      0.01 eV. In the upper panel, only the boundary curves are
      shown. These curves mark the threshold beyond which the two
      states become degenerate, i.e., for $R$ larger than the
      boundary, $E_1 \approx E_2$. At large $R$, these two states
      become degenerate. However, the onset of degeneracy exhibits a
      clear $\theta$-dependence. At the equilibrium bond length $r =
      2.15~a_0$ (black), the dissociation energy is approximately 5.2
      eV, whereas at $r = 1.97~a_0$ (red) and $r = 2.38~a_0$ (blue),
      it increases to about 5.8 eV.}
    \label{sifig:manifold_2}
\end{figure}

\begin{figure}[h!]
    \centering \includegraphics[width=1.0\linewidth]{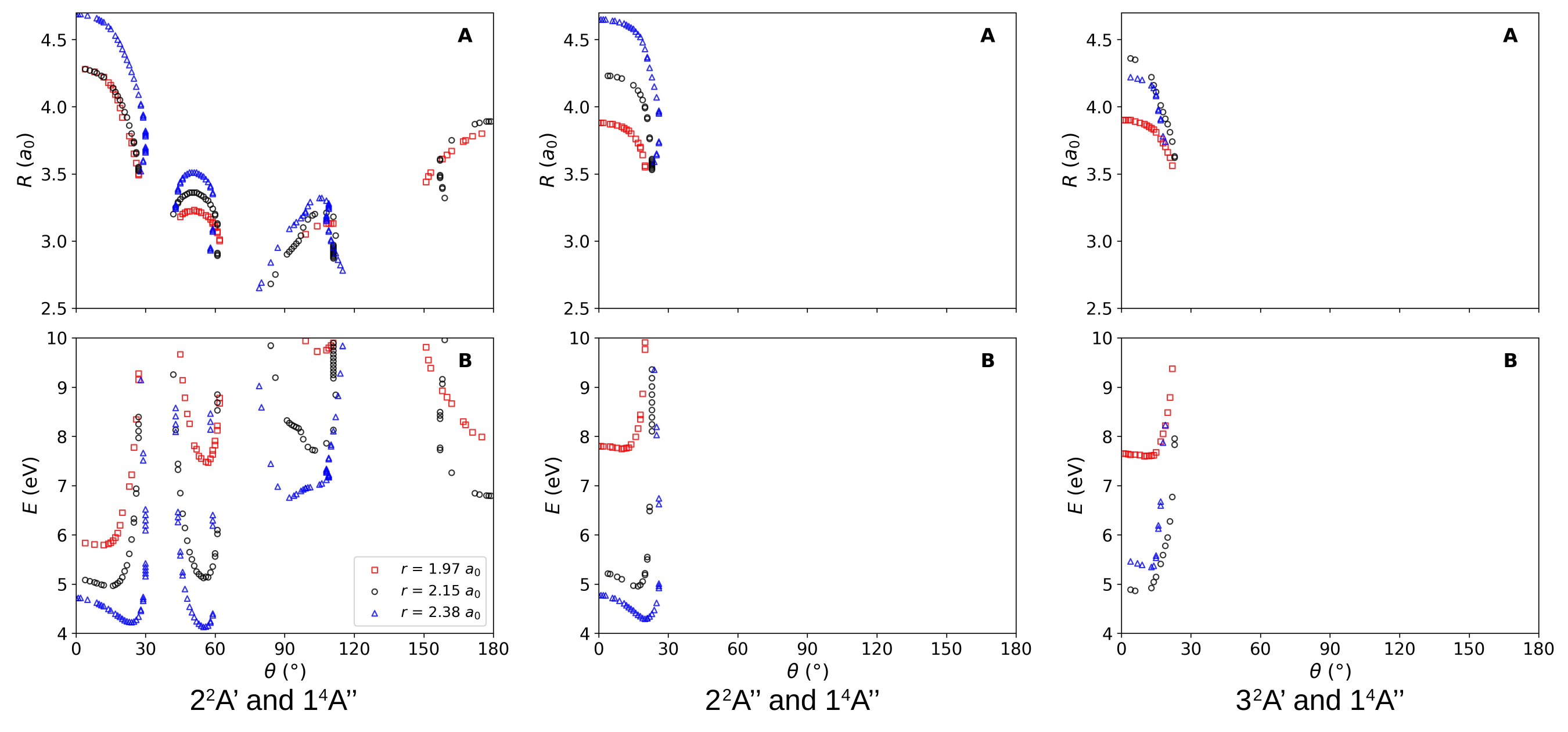}
    \caption{Crossing manifolds of several selected paired of
      electronic states are shown. Each point in the figure
      corresponds to a geometry at which the two states have the same
      energy, with an energy difference smaller than 0.01 eV. A clear
      $\theta$-dependence is observed for the $2 ^2{\rm A}'$ vs $1
      ^4{\rm A}''$ pairs. In contrast, for the $2 ^2{\rm A}''$ vs $1
      ^4{\rm A}''$ and $3 ^2{\rm A}'$ vs $1 ^4{\rm A}''$ pairs, the
      crossings occur only at small $\theta$.}
    \label{sifig:manifold_3}
\end{figure}

\begin{figure}[h!]
    \centering \includegraphics[width=1.0\linewidth]{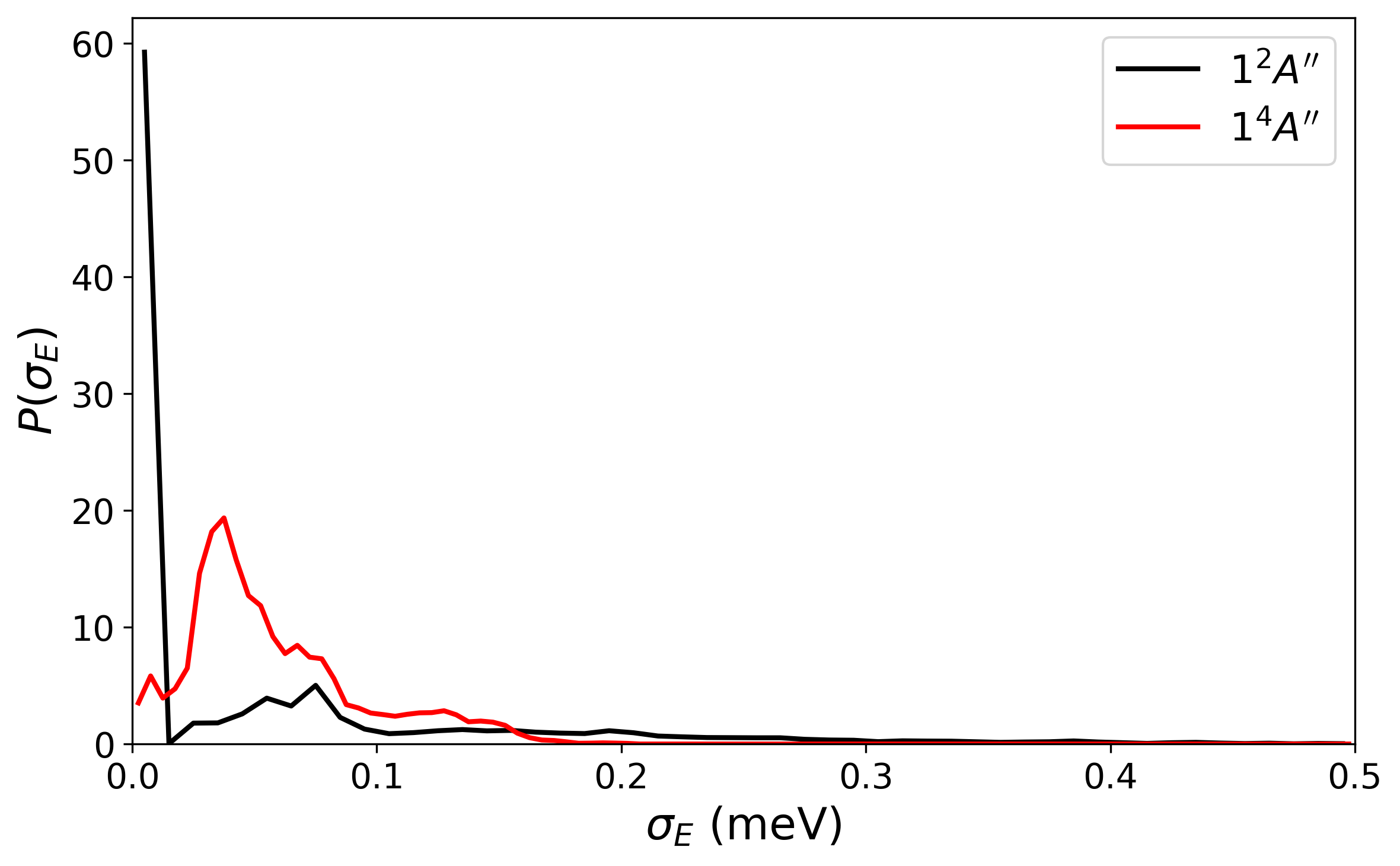}
    \caption{The distributions of the maximum deviation of total
      energy for the $1 ^2{\rm A}''$ (black) and $1 ^4{\rm A}''$ (red)
      states are shown. QCT simulations were carried out using a time
      step of $\Delta t = 0.05$ fs. Energy conservation along each
      trajectory was monitored by evaluating the maximum deviation in
      total energy during propagation. For each trajectory, the
      corresponding maximum deviation, $\sigma_E$, was recorded. Thus,
      for a set of 10000 trajectories, 10000 values of $\sigma_E$ were
      obtained, and their distribution was analyzed using a
      histogram. In all simulations, the energy error never exceeded
      0.5 meV, demonstrating the high accuracy and numerical stability
      of the PES.}
    \label{sifig:cons_hist}
\end{figure}

\begin{figure}[h!]
    \centering \includegraphics[width=1.0\linewidth]{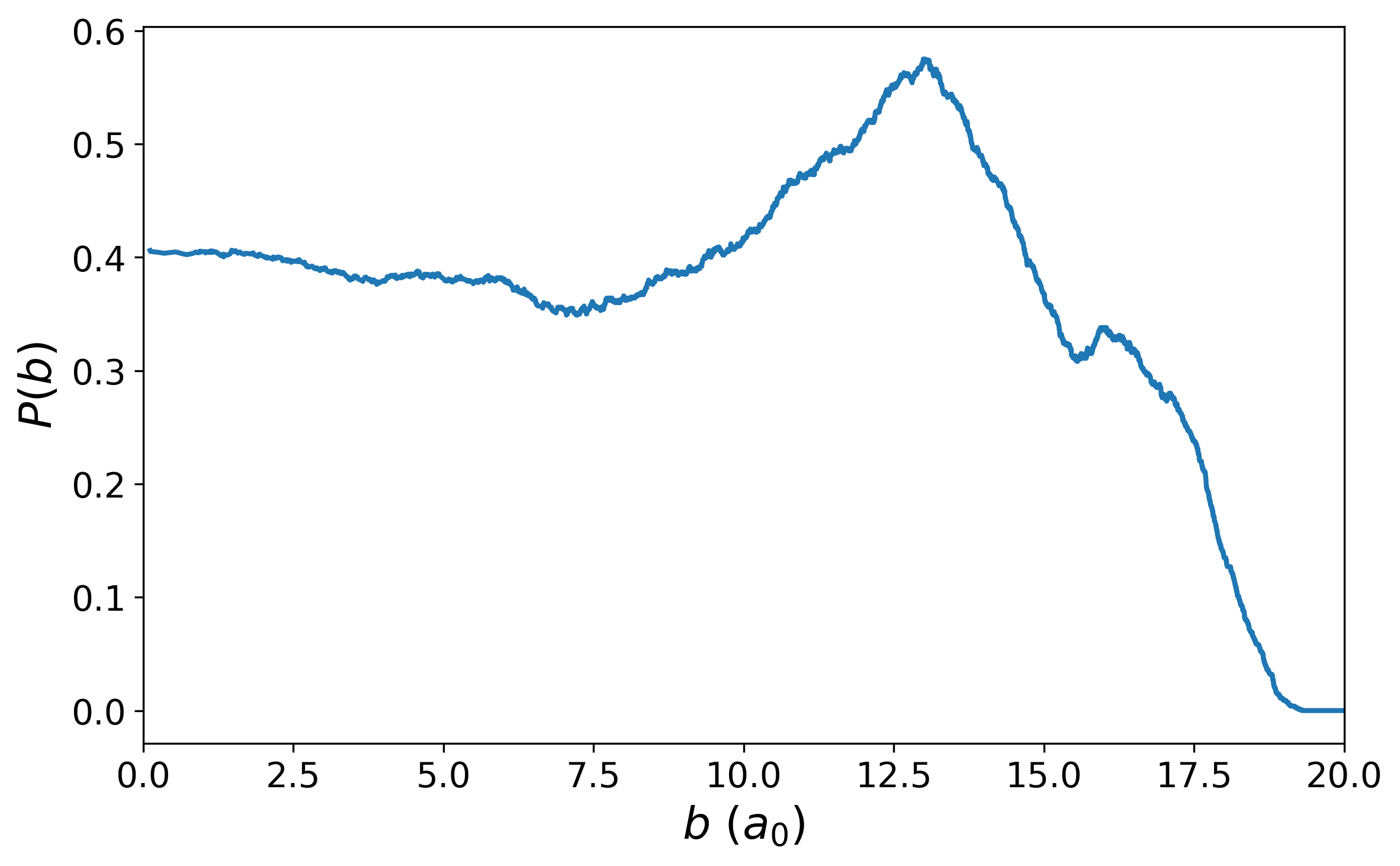}
    \caption{Opacity function from 10000 simulations on the $1 ^2{\rm
        A}''$ ground state RKHS PES, with an initial collision energy
      at 0.05 eV. A complex is considered formed when the total
      distance criterion, $(r_{\mathrm{CO}} + r_{\mathrm{CS}} +
      r_{\mathrm{OS}}) < 15$ a$_0$, is satisfied continuously for
      longer than 1 ps during a simulation. The peak at around $b =
      13~a_0$ is expected. For large $b$, the separation is too great
      for CO and S$^+$ to interact effectively, whereas for very small
      $b$, CO does not have sufficient time to align properly,
      preventing S$^+$ from attaching to the C atom and forming the
      global minimum.}
    \label{sifig:opacity_pb}
\end{figure}

\begin{figure}[h!]
    \centering \includegraphics[width=1.0\linewidth]{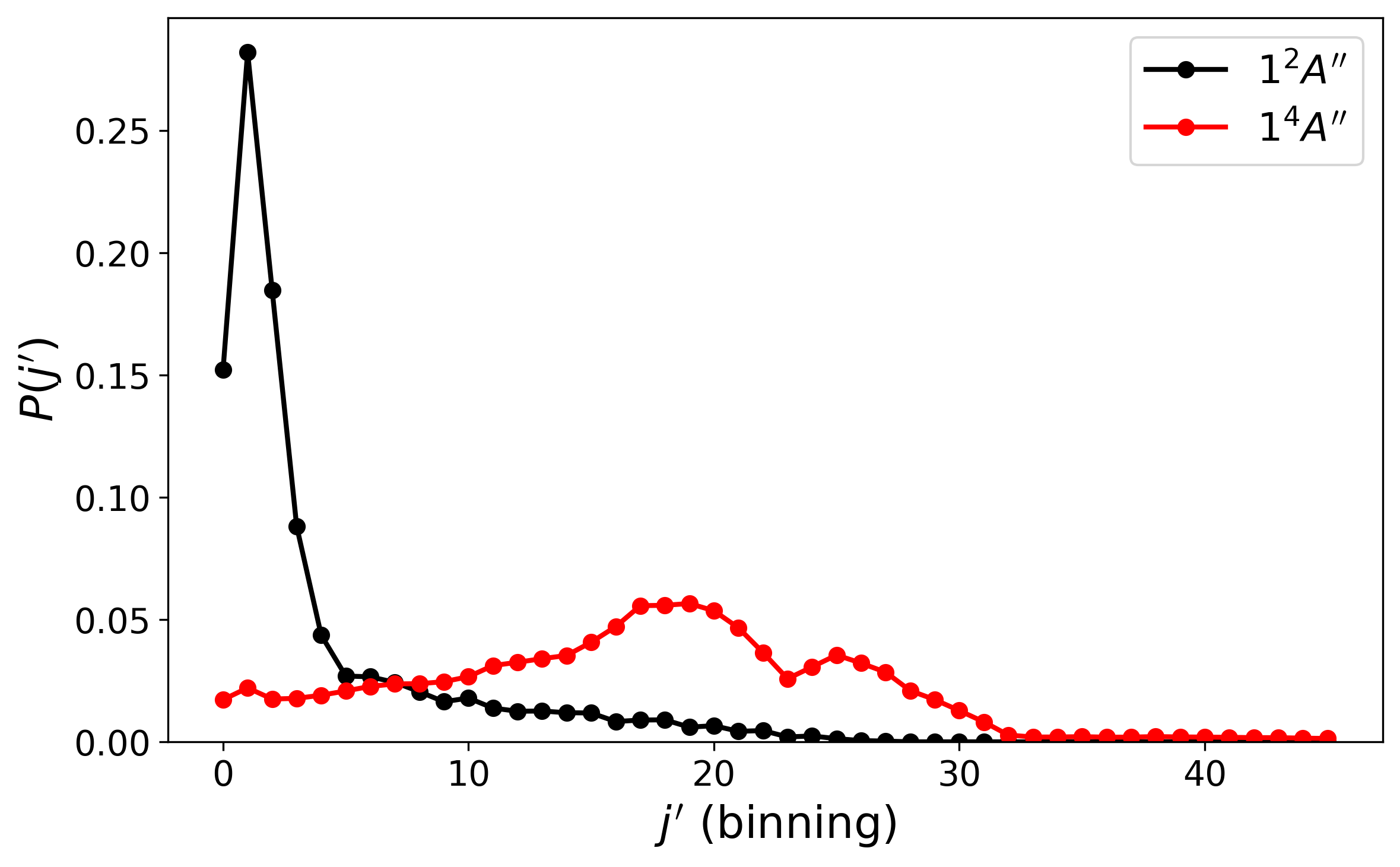}
    \caption{Rotational quantum number distribution from 10000
      independent trajectories for each of the $1 ^2{\rm A}''$ (black)
      and $1 ^4{\rm A}''$ (red) states. For the $1 ^2{\rm A}''$ state,
      the initial collision energy was 0.05~eV. Most trajectories
      dissociate within 100~ps; only 22 exceed this timescale, with
      the longest lasting about 600~ps. In contrast, on the $1 ^4{\rm
        A}''$ PES (red), the initial temperature was set to 5~K
      (corresponding to 0.00043 eV), all the trajectories dissociate
      within 0.02–0.06 ps. More details see Methods section.}
    \label{sifig:j_dis}
\end{figure}

\begin{figure}[h!]
    \centering \includegraphics[width=1.0\linewidth]{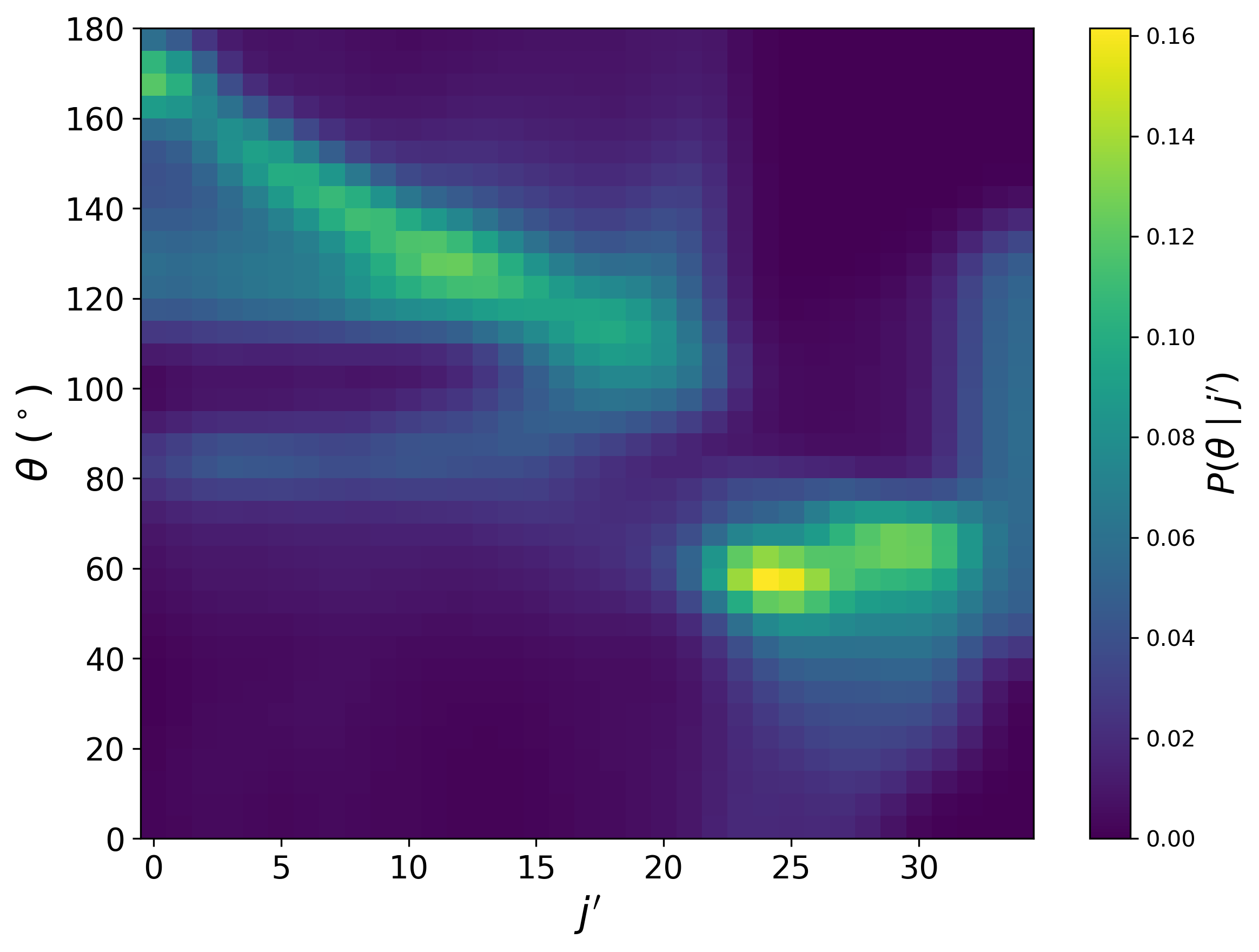}
    \caption{Normalized probability distribution $P(j,\theta)$ of the
      trajectories on the $1 ^4{\rm A}''$ PES, where $j'$ is the
      rotational quantum number of the final state of each trajectory
      and $\theta$ is the the Jacobi angle of the geometries along the
      trajectories. The lower-right region relates to the first
      minimum of the $1 ^4{\rm A}''$ PES, where trajectories tend to
      remain longer and, upon dissociation, result in stronger
      rotational excitation. In contrast, the upper-left region
      relates to the second minimum, which is flatter and leads to
      weaker rotational excitation after dissociation.}
    \label{sifig:j_vs_theta}
\end{figure}

\begin{figure}[h!]
    \centering \includegraphics[width=1.0\linewidth]{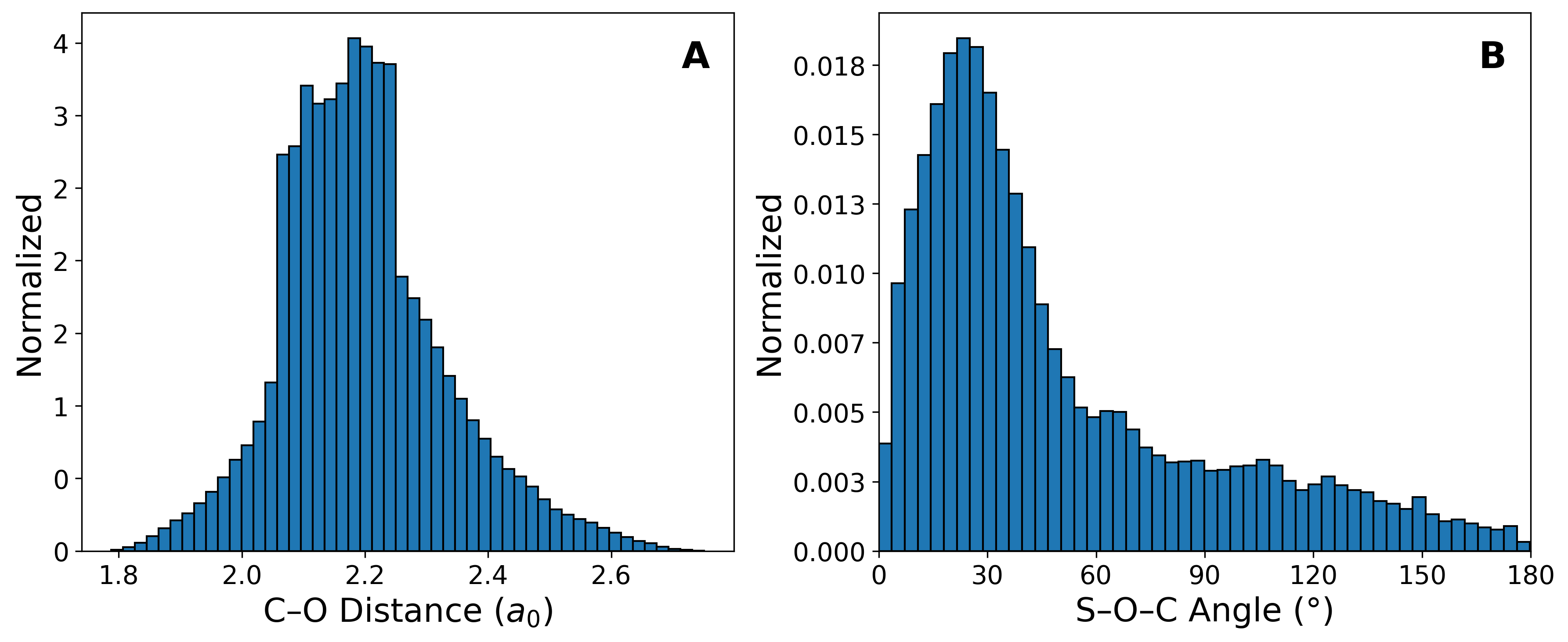}
    \caption{CO bond length (Panel A) and SOC angle (Panel B)
      distributions from 5 independent trajectories for $1 ^2{\rm
        A}''$ ground state. The CO bond length can be compressed or
      elongated to 1.8 or 2.7 ${\rm a}_0$, respectively, which
      corresponds to an energy of about 3 eV, a reasonable range
      considering that the initial energy is approximately 4.0 eV. The
      SOC angle distribution shows that S$^+$ prefers to approach the
      carbon side, consistent with the fact that the carbon atom
      carries a partial negative charge in the CO molecule and with
      the global minimum corresponding to a linear OCS$^+$ structure.}
    \label{sifig:r_co_and_a_soc}
\end{figure}

\clearpage
\bibliography{ref}

@string{cprr={Chem.\ Phys.\ Res.\ Rep.}}

@string{ijqc={Int.\ J.\ Quant.\ Chem.}}

@string{jcp={J.~Chem.\ Phys.}}

@string{jcim={J. Chem. Inf. Model.}}

@string{jpb={J.~Phys.~B:~At.~Mol.~Opt.~Phys.}}

@string{jpca={J.~Phys.\ Chem.~A}}

@string{ma={Math.\ Ann.}}

@string{mp={Mol.\ Phys.}}

@string{pccp={Phys.\ Chem.\ Chem.\ Phys.}}

@article{MM.rkhs:2017,
  title        = {
    Toolkit for the Construction of Reproducing Kernel-based Representations of
    Data: Application to Multidimensional Potential Energy Surfaces
  },
  author       = {Unke, Oliver T. and Meuwly, Markus},
  year         = 2017,
  journal      = jcim,
  volume       = 57,
  number       = 8,
  pages        = {1923--1931}
}

@misc{MOLPRO,
  title        = {MOLPRO, version 2020, a package of ab initio programs},
  author       = {
    H.-J. Werner and P. J. Knowles and G. Knizia and F. R. Manby and M.
    {Sch\"{u}tz} and P. Celani and W. Gy\"orffy and D. Kats and T. Korona and
    R. Lindh and A. Mitrushenkov and G. Rauhut and K. R. Shamasundar and T. B.
    Adler and R. D. Amos and S. J. Bennie and A. Bernhardsson and A. Berning
    and D. L. Cooper and M. J. O. Deegan and A. J. Dobbyn and F. Eckert and E.
    Goll and C. Hampel and A. Hesselmann and G. Hetzer and T. Hrenar and G.
    Jansen and C. K\"oppl and S. J. R. Lee and Y. Liu and A. W. Lloyd and Q. Ma
    and R. A. Mata and A. J. May and S. J. McNicholas and W. Meyer and T. F.
    {Miller III} and M. E. Mura and A. Nicklass and D. P. O'Neill and P.
    Palmieri and D. Peng and K. Pfl\"uger and R. Pitzer and M. Reiher and T.
    Shiozaki and H. Stoll and A. J. Stone and R. Tarroni and T. Thorsteinsson
    and M. Wang and M. Welborn
  },
  year         = 2020,
  address      = {Cardiff, UK}
}

@article{level,
  title        = {LEVEL 6.1},
  author       = {R. J. {Le Roy}},
  year         = 1996,
  journal      = CPRR,
  volume       = {CP-555R},
  pages        = {},
  note         = {}
}

@article{rabitz:1996,
  title        = {A general method for constructing multidimensional
                  molecular potential energy surfaces from ab initio
                  calculations},
  author       = {Tak-San Ho and Herschel Rabitz},
  year         = 1996,
  journal      = jcp,
  volume       = 104,
  pages        = 2584
}

@article{MM.heh2:1999,
  title        = {The potential energy surface and near-dissociation
                  states of He--H$_2^+$},
  author       = {Meuwly, Markus and Hutson, Jeremy M},
  year         = 1999,
  journal      = jcp,
  publisher    = {American Institute of Physics},
  volume       = 110,
  number       = 7,
  pages        = {3418--3427}
}

@article{MM.n3:2024,
  title        = {High-Energy Reaction Dynamics of N$_3$},
  author       = {Wang, JingChun and San Vicente Veliz, Juan Carlos and Meuwly, Markus},
  year         = 2024,
  journal      = jpca,
  publisher    = {ACS Publications},
  volume       = 128,
  number       = 39,
  pages        = {8322--8332}
}

@article{Langhoff:1974,
  title        = {Configuration interaction calculations on the nitrogen molecule},
  author       = {Langhoff, Stephen R. and Davidson, Ernest R.},
  journal      = ijqc,
year= 1974,
  volume       = 8,
  number       = 1,
  pages        = {61--72}
}

@article{Werner:1988,
  title        = {
    An efficient internally contracted multiconfiguration–reference
    configuration interaction method
  },
  author       = {Werner, Hans‐Joachim and Knowles, Peter J.},
  year         = 1988,
  journal      = jcp,
  volume       = 89,
  number       = 9,
  pages        = {5803--5814}
}

@article{Dunning:1989,
  title        = {
    Gaussian basis sets for use in correlated molecular calculations. I. The
    atoms boron through neon and hydrogen
  },
  author       = {Dunning, Thom H., Jr.},
  year         = 1989,
  journal      = jcp,
  volume       = 90,
  number       = 2,
  pages        = {1007--1023}
}

@article{chen:2006,
  title={Dissociation of the OCS$^+$ ion in low-lying electronic states studied using multiconfiguration second-order perturbation theory},
  author={Chen, Bo-Zhen and Chang, Hai-Bo and Huang, Ming-Bao},
  journal=jcp,
  volume={125},
  number={5},
  pages={054310},
  year={2006},
  publisher={AIP Publishing}
}

@article{hirst:2006,
  title={\textit{Ab initio} potential energy surfaces for excited states of the OCS$^+$ molecular ion},
  author={Hirst, David M},
  journal={Mol. Phys.},
  volume={104},
  number={1},
  pages={55--60},
  year={2006},
  publisher={Taylor \& Francis}
}

@article{eland:1973,
  title={Predissociation of N$_2$O$^+$ and COS$^+$ ions studied by photoelectron-photoion coincidence spectroscopy},
  author={Eland, JHD},
  journal={Int. J. Mass Spectrom. Ion Phys.},
  volume={12},
  number={4},
  pages={389--395},
  year={1973},
  publisher={Elsevier}
}

@incollection{truhlar:1979,
	title        = {Atom - Molecule Collision Theory},
	author       = {Truhlar, D. G. and Muckerman, J. T.},
	year         = 1979,
	booktitle    = {Atom--Molecule Collision Theory},
	publisher    = {Springer},
	address      = {New York},
	pages        = {505--566},
	editor       = {Bernstein, R. B.},
}

@book{henriksen2008reaction_dynamics,
	title        = {Theories of Molecular Reaction Dynamics},
	author       = {Henriksen, Niels E. and Hansen, Flemming Y.},
	year         = 2008,
	publisher    = {Oxford University Press},
	address      = {Oxford},
}

@article{koner:2016,
	title        = {State-to-State Dynamics of the ${\rm Ne} +
                  {\rm HeH}^+ (v = 0, j = 0) \rightarrow {\rm NeH}^+
                  (v', j') + {\rm He}$ Reaction },
	author       = {Debasish Koner and Lizandra Barrios and
                  Tom{\'a}s González-Lezana and Aditya N. Panda },
	year         = 2016,
	journal      = jpca,
	volume       = 120,
	number       = 27,
	pages        = {4731--4741},
}

@article{karplus:1965,
	title        = {Exchange Reactions with Activation
                  Energy. I. Simple Barrier Potential for (H, H$_2$) },
	author       = {Karplus, Martin and Porter, Robert N. and Sharma, R. D.},
	year         = 1965,
	journal      = jcp,
	volume       = 43,
	pages        = {3259--3287},
}

@article{MM.cno:2018,
	title        = {The C($^3$P) + NO(X$^2\Pi$) $\rightarrow$
                  O($^3$P) + CN(X$^2\Sigma^+$), N($^2$D)/N($^4$S) +
                  CO(X$^1\Sigma^+$) reaction: Rates, branching ratios,
                  and final states from 15 K to 20 000 K },
	author       = {Koner, Debasish and Bemish, Raymond J. and
                  Meuwly, Markus},
	year         = {2018},
	journal      = jcp,
	volume       = {149},
	number       = {9},
	pages        = {094305},
}

@article{MM.no2:2020,
	title        = {The N($^4$S)+ O$_2$(X$^3\Sigma$) $\rightarrow$
                  O($^3$P)+ NO(X$^2\Pi$) reaction: thermal and
                  vibrational relaxation rates for the 2A$'$, 4A$'$and
                  2A$''$states },
	author       = {San Vicente Veliz, Juan Carlos and Koner,
                  Debasish and Schwilk, Max and Bemish, Raymond J and
                  Meuwly, Markus },
	year         = 2020,
	journal      = pccp,
	publisher    = {Royal Society of Chemistry},
	volume       = 22,
	number       = 7,
	pages        = {3927--3939},
}

@article{MM.co2:2021,
	title        = {The C($^3$P)+ O$_2$($^3 \Sigma_{\rm g}$)
                  $\leftrightarrow$ CO$_2$ $\leftrightarrow$ CO($^1
                  \Sigma^+$) + O($^1$D)/O($^3$P) Reaction: Thermal and
                  Vibrational Relaxation Rates from 15 K to 20000 K },
	author       = {Veliz, Juan Carlos San Vicente and Koner,
                  Debasish and Schwilk, Max and Bemish, Raymond J and
                  Meuwly, Markus },
	year         = 2021,
	journal      = pccp,
	publisher    = {Royal Society of Chemistry},
	volume       = 23,
	number       = 19,
	pages        = {11251--11263},
}

@article{ploenes2021novel,
  title={A novel crossed-molecular-beam experiment for investigating
                  reactions of state-and conformationally selected
                  strong-field-seeking molecules},
  author={Ploenes, L and Stra{\v{n}}{\'a}k, P and Gao, H and
                  K{\"u}pper, J and Willitsch, S},
  journal=mp,
  volume={119},
  number={17-18},
  pages={e1965234},
  year={2021},
  publisher={Taylor \& Francis}
}

@article{lomas25a,
  title={A multimass velocity-map and covariance-map imaging study of the dissociative electron ionisation dynamics of carbonyl sulphide (OCS)},
  author={Lomas, James and Heathcote, David and De Matos Loja, Alexandre and Mile{\v{s}}evi{\'c}, Dennis and Robertson, Patrick and Paterson, Martin J and Vallance, Claire},
  journal=jpb,
  volume={58},
  number={1},
  pages={015202},
  year={2025},
  publisher={IOP Publishing}
}

@article{MM.o3:2025,
  title        = {High-energy reaction dynamics of {O$_3$}},
  author       = {Wang, Jingchun and San Vicente Veliz, Juan Carlos
                  and Upadhyay, Meenu and Meuwly, Markus},
  journal      = jcp,
  volume       = {163},
  number       = {7},
  pages        = {074109},
  year         = {2025},
}

@article{ma21a,
  title={Inverse modelling of carbonyl sulfide: implementation, evaluation and implications for the global budget},
  author={Ma, Jin and Kooijmans, Linda MJ and Cho, Ara and Montzka, Stephen A and Glatthor, Norbert and Worden, John R and Kuai, Le and Atlas, Elliot L and Krol, Maarten C},
  journal={Atmos. Chem. Phys.},
  volume={21},
  number={5},
  pages={3507--3529},
  year={2021},
  publisher={Copernicus GmbH}
}

@article{matthews87a,
  title={Observations of OCS and a search for OC$_3$S in the interstellar medium},
  author={Matthews, HE and MacLeod, JM and Broten, NW and Madden, SC and Friberg, P},
  journal={Astrophys. J.},
  volume={315},
  pages={646--653},
  year={1987}
}

@article{wang24a,
  title   = {High-Resolution Imaging Study on Photodissociation of
                  {OCS}$^{+}$ [{A}$^{2}\Pi_{\Omega=1/2,3/2}$
                  ($\nu_{1}$ 0 $\nu_{3}$)]},
  author  = {Wang, Yaling and Zhao, Yunfan and Luo, Chang and Zhang, Ning and Wang, Wenxin and Hu, Liru and Yuan, Daofu and Wang, Xingan},
  journal = jpca,
  year    = {2024},
}

@article{wang24b,
  title={Vibrational state-specific nonadiabatic photodissociation
                  dynamics of OCS$^+$ via A$^2\Pi_{1/2}$ ($\nu_1$ 0
                  $\nu_3$) states},
  author={Wang, Yaling and Zhao, Yunfan and Zhang, Ning and Wang, Wenxin and Hu, Liru and Luo, Chang and Yuan, Daofu and Zhou, Xiaoguo and Parker, David H and Yang, Xueming and others},
  journal=jcp,
  volume={160},
  number={8},
  pages={084301},
  year={2024},
  publisher={AIP Publishing}
}

@article{chang05a,
  title={Imaging the Mode-Selected Predissociation of OCS$^+~[(v 1 v 2 v 3) \tilde{B}~^2\Sigma^+]$},
  author={Chang, Chushuan and Luo, Chu-Yung and Liu, Kopin},
  journal={J. Phys. Chem. A},
  volume={109},
  number={6},
  pages={1022--1025},
  year={2005},
  publisher={ACS Publications}
}

@article{wiese19a,
  title={Strong-field photoelectron momentum imaging of OCS at finely resolved incident intensities},
  author={Joss Wiese and Jean-Francois Olivieri and Andrea Trabattoni and Sebastian Trippel and Jochen K\"upper},
  journal={New J. Phys.},
  volume={21},
  pages={083011},
  year={2019},
  }

@article{holland16a,
  title={A study of the valence shell absolute photoabsorption, photoionisationand photodissociation cross sections and the photoionisation quantumefficiency of carbonyl sulphide},
  author={D.M.P. Holland and D.A. Shaw},
  journal={Chem. Phys.},
  volume={479},
  pages={151-159},
  year={2016},
  }

@article{ramadhan16a,
  title={Ultrafast molecular dynamics of dissociative ionization in OCS probed by soft x-ray synchrotron radiation},
  author={Ali Ramadhan and Benji Wales and Reza Karimi and Isabelle Gauthier and Michael MacDonald and Lucia Zuin and Joe Sanderson},
  journal={J. Phys. B: At. Mol. Opt. Phys.},
  volume={49},
  pages={215602},
  year={2016},
  }

@article{kishimoto03a,
	author = {Kishimoto, Naoki and Horio, Takuya and Maeda, Satoshi and Ohno, Koichi},
	journal = {Chem. Phys. Lett.},
	title = {Collision-energy-resolved Penning ionization electron spectroscopy of OCS with He(2$^3$S) metastable atoms},
	pages = {332},
	volume = {379},
	year = {2003}}

@article{horio06a,
  title={Anisotropic interaction and stereoreactivity in a chemi-ionization process of OCS by Collision with He*(2$^3$S) Metastable Atoms},
  author={Horio, Takuya and Maeda, Satoshi and Kishimoto, Naoki and Ohno, Koichi},
  journal=jpca,
  volume={110},
  number={38},
  pages={11010--11017},
  year={2006},
  publisher={ACS Publications}
}

@article{wang88a,
  title={High resolution UV photoelectron spectroscopy of CO$^+_2$,
                  COS$^+$ and CS$^+_2$ using supersonic molecular
                  beams},
  author={Wang, Lai-Sheng and Reutt, JE and Lee, YT and Shirley, DA},
  journal={J. Electron. Spectrosc. Rel. Phen.},
  volume={47},
  pages={167--186},
  year={1988},
  publisher={Elsevier}
}

@article{dong2011loss,
  title={O-loss photodissociation of the OCS$^+$ ion in the low-lying
                  electronic states studied using multiconfiguration
                  second-order perturbation theory},
  author={Dong, Hua and Chen, Bo-Zhen and Huang, Ming-Bao and Chang,
                  Hai-Bo},
  journal={Int. J. Quantum Chem.},
  volume={111},
  number={14},
  pages={3578--3587},
  year={2011},
  publisher={Wiley Online Library}
}

\end{document}